\documentclass[aps,prd,preprint]{revtex4-1}
\usepackage[utf8]{inputenc}
\usepackage[pdftex]{graphicx}
\usepackage{amsmath,amssymb}
\usepackage{mathtools}
\usepackage{physics}
\usepackage{hyperref}
\usepackage{duckuments}
\hypersetup{
    colorlinks,
    citecolor=blue,
    filecolor=blue,
    linkcolor=blue,
    urlcolor=blue
}

\begin{document}
\title{Exploring the DGLAP resummation in the JIMWLK Hamiltonian}
\date{\today}
\author{N\'estor Armesto$^a$, Alex Kovner$^{b,c,d}$ and  V\'{\i}ctor L\'opez-Pardo$^a$}

\affiliation{$^a$ Instituto Galego de F\'{\i}sica de Altas Enerx\'{\i}as IGFAE, Universidade de Santiago de Compostela, 15782 Santiago de Compostela, Galicia-Spain\\$^b$ Physics Department, University of Connecticut, Storrs, CT 06269, USA\\
$^c$ ExtreMe Matter Institute EMMI,
GSI Helmholtzzentrum fuer Schwerionenforschung GmbH,
Planckstrasse 1,
64291 Darmstadt,
Germany\\
$^d$ Theoretical Physics Department, CERN, CH-1211 Geneve 23, Switzerland}
\preprint{CERN-TH-2025-023}
\begin{abstract}
{We explore the recently derived equation that resums DGLAP corrections to the JIMWLK Hamiltonian in the simplified setting of the $SU(2)$ gauge theory. We solve the equation numerically for the scattering matrix of a dressed gluon for a particular initial condition, that corresponds to a dipole initial state. {\color{black}We evolve  the $S$-matrix of a single dressed gluon from the scale $Q_P$, which is the inverse color correlation length in the projectile to $Q\gg Q_P$ which corresponds to the hard resolution scale provided by the target. As expected, $S$} ceases to be unitary if evolved to significant {\color{black}$\ln Q^2/Q_P^2$}. Our numerical results indicate an interesting universal (independent of the coupling constant) pattern for this deviation from unitarity.
}
\end{abstract}
\maketitle
\tableofcontents

\section{Introduction}
 In recent years there is a significant effort~\cite{Balitsky:2015qba,Balitsky:2016dgz,Balitsky:2023hmh,Xiao:2017yya,Boussarie:2021wkn,Boussarie:2023xun,Mukherjee:2023snp,Caucal:2024bae,Duan:2024qck,Duan:2024qev} to understand how to controllably connect the saturation regime in hadronic collisions at high energy with the intermediate-$x$ physics, such as Dokshitzer-Gribov-Lipatov-Altarelli-Parisi (DGLAP)~\cite{Gribov:1972ri,Gribov:1972rt,Altarelli:1977zs,Dokshitzer:1977sg} or Collins-Soper-Sterman (CSS)~\cite{Collins:1981uk,Collins:1984kg,Collins:1987pm,Collins:1989gx} evolution. An initial impetus to this effort was given by the calculation of the next-to-leading order (NLO) corrections~\cite{Fadin:1998py,Ciafaloni:1998gs} to the Balitsky-Fadian-Kuraev-Lipatov (BFKL) equation~\cite{Kuraev:1977fs,Balitsky:1978ic,Lipatov:1985uk}, which rendered the high energy evolution unstable~\cite{Ross:1998xw,Kovchegov:1998ae,Armesto:1998gt} in the absence of a resummation of large transverse logarithms. In addition, this question is becoming more urgent in view of the approval of the Electron Ion Collider (EIC)~\cite{Accardi:2012qut,AbdulKhalek:2021gbh}, as the energies available at the EIC are not going to be high enough so that the BFKL type of physics, and physics of saturation, is not expected to be cleanly separable from the parton model-like Quantum Chromodynamics (QCD).

 While resummation of transverse logarithms in the BFKL framework has been addressed right from the appearance of the NLO BFKL calculation~\cite{Salam:1998tj,Ciafaloni:1998iv,Ciafaloni:1999yw,Ciafaloni:1999au,Ciafaloni:2003ek,Ciafaloni:2003rd,Altarelli:2000mh,Altarelli:2001ji,Altarelli:2005ni,Altarelli:2008aj,Kutak:2004ym,SabioVera:2005tiv,Motyka:2009gi}, a similar effort in the saturation domain within the Balitsky-Kovchegov (BK) or Jalilian-Marian--Iancu--Mclerran--Leonidov--Kovner (JIMWLK) evolution~\cite{Balitsky:1995ub,Balitsky:1998kc,Balitsky:1998ya,Jalilian-Marian:1997qno,Jalilian-Marian:1997jhx,Jalilian-Marian:1997ubg,Kovchegov:1999ua,Kovner:1999bj,Kovner:2000pt,Weigert:2000gi,Iancu:2000hn,Iancu:2001ad,Ferreiro:2001qy,Balitsky:2007feb,Balitsky:2013fea,Grabovsky:2013mba,Kovner:2013ona,Kovner:2014lca,Lublinsky:2016meo}, is a much more recent development~\cite{Beuf:2014uia,Iancu:2015joa,Iancu:2015vea,Ducloue:2019ezk,Lappi:2016fmu,Kovner:2023vsy}. Very recently it was realized that a certain set of corrections to the NLO BK equation has its origin in DGLAP evolution. The terms in question have been previously attributed to the running of the QCD coupling constant \cite{Balitsky:2006wa,Kovchegov:2006vj}, since they are proportional to the QCD $\beta$-function. However, as shown in~\cite{Kovner:2023vsy}, the correct way of resumming this set of logarithms is into DGLAP splittings of the gluons of the projectile which are not accounted for in the double logarithmic approximation. The contribution of these terms is also proportional to the $\beta$-function, but they have a very different physical origin than the running of $\alpha_s$. 
 
 {\color{black} It is indeed completely natural to expect the appearance of DGLAP logarithms in NLO JIMWLK equation, since at NLO one has to encounter the part of the gluon splitting function which is not included in the eikonal vertex. These are precisely the logarithms resummed by the equation written in~\cite{Kovner:2023vsy}. Physically these are important when the scattering process involves a projectile and a target with very different color correlations lengths in the transverse plane. For such processes the splittings in the projectile wave function that produce gluons with high transverse momenta, up to the inverse correlation length of the target, have to be accounted for. This is taken care of by the resummation of~\cite{Kovner:2023vsy}.}

The resummation of DGLAP splittings derived in~\cite{Kovner:2023vsy} is achieved by defining the $S$-matrix of a dressed gluon state with resolution $Q$, i.e., $\mathbb{S}_Q$. This $S$-matrix satisfies the analog of the DGLAP equation~\footnote{One should keep in mind that~\eqref{DGLAP} is not the full DGLAP equation, as it only resumms splittings which are not already resummed in the double logarithmic regime. Those latter splittings are already present in the JIMWLK evolution. As a result of this,~\eqref{DGLAP} is somewhat peculiar in that it actually cannot be interpreted in terms of splitting probabilities, as it {\it subtracts} probabilities which are taken to be too large in JIMWLK evolution, which approximates the DGLAP splitting function by its low $x$ asymptotics in the full range of $x$.}
\begin{equation}\label{DGLAP}
   \frac{\partial}{\partial \ln Q^2} \mathbb{S}_Q^{ab}(\mathbf{z}) = -
    \frac{\alpha_s\beta^g_0}{4\pi} 
    \left[ \mathbb{S}_Q^{ab}(\mathbf{z})-\frac{1}{N_c}\int \frac{d\phi}{2\pi}
        \left( \mathbb{D}_Q^{ab}(\mathbf{z} + \frac{1}{2} Q^{-1} \mathbf{e}_r, \mathbf{z} -\frac{1}{2} Q^{-1} \mathbf{e}_r)   \right) \right],
\end{equation}
where
\begin{align}
    \mathbb{D}_Q^{ab}(\mathbf{z}_1, \mathbf{z}_2) = 
    {\rm Tr}[ T^a  \mathbb{S}_Q(\mathbf{z}_1)  T^b \mathbb{S}_Q^+(\mathbf{z}_2)]\,,
\end{align}
$\mathbf{e}_r$ is the unit vector in the radial direction with $\mathbf{z}$ chosen as origin, $\phi$ is the corresponding polar angle, and $\beta_0^g=11 N_c/3$.

 At $Q=Q_T$, where $Q_T$ is the inverse correlation scale of color fields in the target, one has $\mathbb{S}_{Q_T}=S$, where $S$ is the eikonal scattering matrix of the bare gluon. Eq.~\eqref{DGLAP} has to be integrated with this initial condition down to $Q=Q_P$, where $Q_P$ is the analogous scale in the projectile. The resummation is only necessary when $Q_T\gg Q_P$, or more quantitatively when $\alpha_s\ln\frac{Q_T^2}{Q_P^2}\sim 1$. Eq.~\eqref{DGLAP} was solved approximately in~\cite{Kovner:2023vsy} in two extreme limits -- of dilute and dense target.

 The present paper is dedicated to a more detailed study of~\eqref{DGLAP}. We are particularly interested in the question how far does the evolution takes $\mathbb{S}_Q$ from a unitary matrix, which it is at $Q=Q_T$. 
 It is important to realize that $\mathbb{S}_Q$ is not required to be unitary, since it is a scattering matrix of a dressed gluon state. For {\color{black}$Q>Q_T$} the internal structure of the gluon is not resolved by the target. 
 {\color{black}Therefore at these high values of the hard scale $\mathbb{S}_Q$ is a unitary $3\times 3$ matrix. This is the case }  since the only effect of scattering in the eikonal approximation is to rotate the color. In other words, a pointlike gluon is an eigenstate of the eikonal scattering matrix (up to color rotation), and the unitarity of the full quantum $S$-matrix operator therefore requires unitarity of the eikonal scattering matrix of a pointlike gluon.
 
 On the other hand a dressed gluon whose structure is resolved by the target, i.e., for {\color{black}$Q\le Q_T$}, is not an eigenstate of the eikonal scattering matrix. In particular, it can radiate while scattering via decoherence of its constituents. The physical unitarity condition involves all possible final states in the scattering process, and thus more states than just a single dressed gluon.  {\color{black} The dressed gluon does not propagate through the target coherently. Since it is an extended state with the transverse size greater than the resolution scale of the target, its components in principle scatter with different phases. This decoherence physically is equivalent to emission, or presence  of additional gluons in the final state. On the other hand, $\mathbb{S}_Q$ is an element of the full $S$-matrix between single (dressed) gluon initial and final states. It is thus a} truncated scattering matrix defined only in a {\bf subset} of states connected by scattering. It, therefore, by itself is not required to be unitary. 
The deviation of $\mathbb{S}_Q$ from unitarity is thus an indirect measure of the importance of radiation by a dressed gluon in an eikonal scattering.

 We are not going to study  equation~\eqref{DGLAP} in full generality. Our goal is rather modest, and it is to get an idea of how the evolution affects $S$. Consequently, we are going to study a simplified case of the $SU(2)$ gauge group. For $SU(2)$ we can parametrize the matrix $\mathbb{S}_Q$ in a rather simple way which involves only a small number of degrees of freedom. 
 
 In addition, we limit ourselves to a dilute approximation, that is we assume that throughout the evolution $\mathbb{S}_Q$ remains close to unity. This approximation was studied in~\cite{Kovner:2023vsy}. However here we go beyond~\cite{Kovner:2023vsy} and keep second order corrections to $\mathbb{S}_Q$ which allows us to probe the deviations of $\mathbb{S}_Q$ from a unitary matrix.

 {\color{black} Here one has to keep in mind two important points. First, note that we are using dilute approximation in the situation where there is big hierarchy between the transverse scales in the target and the projectile, $Q_T\gg Q_P$. One could wonder if in this regime saturation effects have to be important, which would invalidate the use of dilute approximation. However it is important to realize that in our discussion $Q_T$ and $Q_P$ are not the values of "saturation momentum", but rather the inverse correlation length. Thus one can perfectly well have an object with large value of $Q_T$, i.e., short correlation length, which is nevertheless dilute. The prototype of such an object is a single small dipole with the size $r^2\ll \Lambda_{QCD}^{-2}$. It is obviously dilute, but the correlation length, which is of order of its size, is nevertheless small. The dilute approximation discussed in the present paper is valid as long as neither the projectile nor the target is dense and the projectile is perturbative. A good example is a dipole-dipole scattering, where both dipoles are small on the strong interaction scale $\Lambda_{QCD}$, and the target dipole is much smaller than the projectile dipole.
 
 Secondly, one could ask why do we need to go to second order in deviation of $\mathbb{S}_Q$ from unity in order to study its deviations from unitarity. As was shown in~\cite{Kovner:2023vsy}, and will be recapitulated below, at first order, the only relevant degree of freedom in $\mathbb{S}_Q$ is the adjoint representation, albeit modified by the resummation. With only an adjoint contribution $\mathbb{S}_Q$ always remains unitarity. It is only at second order that the deviation from unitarity can arise. This is the reason we are including these corrections in the current analysis.}

 Our approach in this paper will be numerical. In Section 2 we write out explicitly the evolution equation for components of $\mathbb{S}_Q$ in the $SU(2)$ case. In Section 3 we derive the equation in the dilute limit keeping terms of order $\alpha_s^2$. In Section 4 we discuss a particular initial condition corresponding to a single dipole target. Finally, in Section 5 we present our numerical results followed by a short discussion in Section 6.

\section{DGLAP equation for  $\mathrm{SU}(2)$}
To facilitate the numerical study of \eqref{DGLAP} we first have to choose 
a convenient parametrization of the matrix $\mathbb{S}_Q$. 
{\color{black}The $S$-matrix of a bare gluon is a unitary $3\times 3$ matrix, since the gluon belongs to an adjoint (vector) representation of $SU(2)$.
This $S$-matrix can be parametrized as
\begin{equation}\label{sglue}
S(\mathbf{x})_{bc}=[e^{iT^a\lambda^a(\mathbf{x})}]_{bc}=1+\epsilon_{abc}\lambda^a(\mathbf{x})+[\epsilon^a\epsilon^d]_{bc}\lambda^a(\mathbf{x})\lambda^d(\mathbf{x})+...
\end{equation}
where we used for the generators of $SU(2)$ in the adjoint representation
\begin{equation}
    T^a_{bc}=-i\epsilon_{abc}\,.
    \end{equation}
}
At an arbitrary value of $Q$ the matrix $\mathbb{S}_Q(\mathbf{x})$ does not have to be unitary. It is nevertheless a real $3\times 3$ matrix whose indices can be thought of as transforming under $SU(2)$ rotations. We can therefore decompose it in terms of the representations of the $SU(2)$ group as follows:
\begin{equation}\label{dec}
\mathbb{S}^{ab}=A\delta_{ab}+\lambda^c\epsilon_{abc}-2B^{ab} ,
\end{equation}
where $B^{ab}=B^{ba}$ and $\mathrm{Tr}\,B=0$. Here $A$ is a rotational scalar, $\lambda^a$ is a vector and $B^{ab}$ belongs to the spin 2 symmetric tensor representation.

With the decomposition \eqref{dec}, after some straightforward algebra we can calculate
\begin{eqnarray}
\mathrm{Tr}\,[T^a\mathbb{S}_1T^b\mathbb{S}_2^\dagger]&=&\delta^{ab}
    \left[2A_1A_2+\frac{2}{3}\lambda_1\cdot\lambda_2-\frac{4}{3}Tr[B_1B_2]\right]\\
&+&\epsilon^s_{ab}\left[A_1\lambda_2^s+A_2\lambda_1^s-2(\lambda_1\cdot B_2)^s-2(\lambda_2\cdot B_1)^s\right]\nonumber\\
&+&\left[\lambda_1^a\lambda_2^b+\lambda_1^b\lambda_2^a-\frac{2}{3}\delta^{ab}\lambda_1\cdot\lambda_2\right]
+2A_1B_2^{ab}+2A_2B_1^{ab}\nonumber \\ 
&+&4\left[(B_1B_2)^{ab}+(B_2B_1)^{ab}-\frac{2}{3}Tr(B_1B_2)\right].\nonumber
\end{eqnarray}
Here, the index $i=1,2$ stands for the values of transverse coordinates $\mathbf{z}_1, \ \mathbf{z}_2$.
To arrive at this we used the following simple identities:
\begin{eqnarray}
    &&\mathrm{Tr}\,[\epsilon^a\epsilon^b\epsilon^c]=\epsilon^{abc}=\epsilon^a_{bc},\\
    &&\epsilon^a_{fc}\epsilon^b_{de}=\delta^{ab}\delta^{fd}\delta^{ce}-\delta^{ab}\delta^{fe}\delta^{cd}-\delta^{ad}\delta^{fb}\delta^{ce}+\delta^{ad}\delta^{fe}\delta^{cb}-\delta^{ae}\delta^{fd}\delta^{cb}+\delta^{ae}\delta^{ae}\delta^{fb}\delta^{cd}\nonumber,\\
   &&[\epsilon^a\epsilon^b]_{ef} =\delta^{af}\delta^{eb}-\delta^{ab}\delta^{ef}.\nonumber
\end{eqnarray}

Eq.~\eqref{DGLAP} can then be written as a set of equations for functions $A$, $\lambda^a$ and $B^{ab}$.
Introducing the explicit $Q$ dependence, we obtain

\begin{eqnarray}
    \frac{\partial}{\partial\ln Q^2}A_Q(\mathbf{z})&=&-\frac{\alpha_s\beta^g_0}{4\pi}\left[A_Q(\mathbf{z})+\frac{1}{2}\int\frac{d\phi}{2\pi}\Big(-2A_Q(\mathbf{z}_1)A_Q(\mathbf{z}_2)\right.\\&&\hskip 1cm \left.-\frac{2}{3}\lambda_Q^c(\mathbf{z}_1)\lambda_Q^c(\mathbf{z}_2)+\frac{4}{3}B_Q^{pq}(\mathbf{z}_1)B_Q^{qp}(\mathbf{z}_2)\Big)\right],\nonumber
\end{eqnarray}
\begin{eqnarray}
    \frac{\partial}{\partial\ln Q^2}\lambda_Q^c(\mathbf{z})&=&-\frac{\alpha_s\beta^g_0}{4\pi}\left[\lambda_Q^c(\mathbf{z})-\frac{1}{2}\int\frac{d\phi}{2\pi}\Big(A_Q(\mathbf{z}_1)\lambda_Q^c(\mathbf{z}_2)+\lambda_Q^c(\mathbf{z}_1)A_Q(\mathbf{z}_2)\right.
    \\&&\hskip 1cm\left.-2\lambda_Q^d(\mathbf{z}_1)B_Q^{dc}(\mathbf{z}_2)-2B_Q^{cd}(\mathbf{z}_1)\lambda_Q^d(\mathbf{z}_2)\Big)\right],\nonumber
\end{eqnarray}
\begin{multline}
    \frac{\partial}{\partial\ln Q^2}B_Q^{cd}(\mathbf{z})=-\frac{\alpha_s\beta^g_0}{4\pi}\left[B_Q^{cd}(\mathbf{z})+\frac{1}{2}\int{\frac{d\phi}{2\pi}\left(A_Q(\mathbf{z}_1)B_Q^{cd}(\mathbf{z}_2)+B_Q^{cd}(\mathbf{z}_1)A_Q(\mathbf{z}_2)\right.}\right.\\\left.{\phantom{\frac{d\phi}{2\pi}}\left.+\lambda_Q^c(\mathbf{z}_1)\lambda_Q^d(\mathbf{z}_2)-\frac{1}{3}\delta_{cd}\lambda_Q^a(\mathbf{z}_1)\lambda_Q^a(\mathbf{z}_2)\right.}\right.\\\left.{\phantom{\frac{d\phi}{2\pi}}\left.+2B_Q^{dp}(\mathbf{z}_1)B_Q^{pc}(\mathbf{z}_2)+2B_Q^{cp}(\mathbf{z}_1)B_Q^{pd}(\mathbf{z}_2)-\frac{4}{3}\delta_{cd}B_Q^{pq}(\mathbf{z}_1)B_Q^{qp}(\mathbf{z}_2)\right)}\right].
\end{multline}
Our aim now is to study the solutions of these equations. These are coupled nonlinear equations and solving them in full generality is a formidable problem. Instead we will restrict ourselves to the dilute limit in which the equations simplify considerably.

\section{The dilute limit}
In~\cite{Kovner:2023vsy} the equation~\eqref{DGLAP} was studied analytically in the dilute limit, i.e., assuming $\mathbb{S}_Q$ is close to the unit matrix. In this limit, {\color{black} as is clear from the representation \eqref{sglue}, the largest deviation from the identity matrix is given by the adjoint component $\lambda^a$, since all other components of $S$ are of order $\lambda^2$. The same is true for $\mathbb{S}_Q(\mathbf{x})$, since for $\alpha_s\ln Q^2/Q_P^2\sim 1$ the evolution changes the values of $\lambda, \ A$ and $B$ by factors of order unity.  The leading contribution to $\mathbb{S}_Q$} comes therefore from the evolution of $\lambda$ since all other components of $\mathbb{S}$ are of order $\lambda^2$.

Neglecting $1-A$ and $B$ one then obtains a closed equation for $\lambda$~\cite{Kovner:2023vsy}. Here we will not  neglect $1-A$ and $B$ entirely, however we will still make the simplifying assumption that $\lambda$ is small. 
Defining $A_Q=1+\Delta_Q$, in the dilute limit ($\lambda\sim \alpha_s\ll1$) the natural scaling is $\Delta, B\propto\lambda^2$. We can then keep only terms of order $\lambda$ in the equation for $\lambda$, and terms of order $\lambda^2$ in the equations for $\Delta$ and $B$. Under these simplifying assumptions the equations become
\begin{eqnarray}
    \frac{\partial}{\partial\ln Q^2}\Delta_Q(\mathbf{z})&=&-\frac{\alpha_s\beta^g_0}{4\pi}\left[1+\Delta_Q(\mathbf{z})\right.\\&&\hskip 1cm\left.+\frac{1}{2}\int{\frac{d\phi}{2\pi}\left(-2-2\Delta_Q(\mathbf{z}_1)-2\Delta_Q(\mathbf{z}_2)-\frac{2}{3}\lambda_Q^c(\mathbf{z}_1)\lambda_Q^c(\mathbf{z}_2)\right)}\right],\nonumber
\end{eqnarray}
\begin{equation}
    \frac{\partial}{\partial\ln Q^2}\lambda_Q^c(\mathbf{z})=-\frac{\alpha_s\beta^g_0}{4\pi}\left[\lambda_Q^c(\mathbf{z})-\frac{1}{2}\int{\frac{d\phi}{2\pi}\left(\lambda_Q^c(\mathbf{z}_2)+\lambda_Q^c(\mathbf{z}_1)\right)}\right],
\end{equation}
\begin{eqnarray}
    \frac{\partial}{\partial\ln Q^2}B_Q^{cd}(\mathbf{z})&=&-\frac{\alpha_s\beta^g_0}{4\pi}\left[B_Q^{cd}(\mathbf{z})\right.\\&&\hskip 0.5cm\left.+\frac{1}{2}\int{\frac{d\phi}{2\pi}\left(B_Q^{cd}(\mathbf{z}_2)+B_Q^{cd}(\mathbf{z}_1)+\lambda_Q^c(\mathbf{z}_1)\lambda_Q^d(\mathbf{z}_2)-\frac{1}{3}\delta_{cd}\lambda_Q^a(\mathbf{z}_1)\lambda_Q^a(\mathbf{z}_2)\right)}\right].\nonumber
\end{eqnarray}
The simplifying power of the approximation above is that the equation for $\lambda$ remains decoupled from $\Delta$ and $B$, and therefore can be solved independently. The terms that involve $\lambda$ then appear as source (inhomogeneous) terms in the equations for $\Delta$ and $B$, which still remain decoupled from each other. 

In this way all the equations are linear, albeit some of them are inhomogeneous. Thus in this approximation we are able to solve the equations, at least formally, in closed form and express the solutions in terms of initial conditions. This is what we are going to do first. Once we have the solutions we will choose a physical initial condition, and will find numerically $\mathbb{S}_Q$ that corresponds to this initial condition.

\subsection{Solving for $\lambda$}

We start by solving the equation for $\lambda$:
\begin{equation}
    \frac{\partial}{\partial\ln Q^2}\lambda_Q^c(\mathbf{z})=-\frac{\alpha_s\beta^g_0}{4\pi}\left[\lambda_Q^c(\mathbf{z})-\frac{1}{2}\int{\frac{d\phi}{2\pi}\left(\lambda_Q^c(\mathbf{z}+Q^{-1}\mathbf{e}_r/2)+\lambda_Q^c(\mathbf{z}-Q^{-1}\mathbf{e}_r/2)\right)}\right].
\end{equation}
Transforming this equation into momentum space we get
\begin{eqnarray}
    \frac{\partial}{\partial\ln Q^2}\lambda_Q^c(\mathbf{p})&=&-\frac{\alpha_s\beta^g_0}{4\pi}\left[\lambda_Q^c(\mathbf{p})-\frac{1}{2}\int{\frac{d\phi}{2\pi}\left(e^{i\frac{\mathbf{p}\cdot\mathbf{e}_r}{2Q}}\lambda_Q^c(\mathbf{p})+e^{-i\frac{\mathbf{p}\cdot\mathbf{e}_r}{2Q}}\lambda_Q^c(\mathbf{p})\right)}\right]\\
    &=&-\frac{\alpha_s\beta^g_0}{4\pi}\left[\lambda_Q^c(\mathbf{p})-J_0\left(\frac{p}{2Q}\right)\lambda_Q^c(\mathbf{p})\right],\nonumber
\end{eqnarray}
{\color{black}where we use the notation $p\equiv |\mathbf{p}|$.}

Defining
\begin{equation}
    R(p,Q)=\frac{\alpha_s\beta^g_0}{4\pi}\left[J_0\left(\frac{p}{2Q}\right)-1\right],
\end{equation}
we have
\begin{equation}
    \label{eq:sollambda}
\lambda_Q^c(\mathbf{p})=\exp\left[-\int_Q^{Q_T}{\frac{dQ^{\prime 2}}{Q^{\prime 2}}R(p,Q^\prime)}\right]\lambda_{Q_T}^c(\mathbf{p}).
\end{equation}
This is, of course, the same solution as found in~\cite{Kovner:2023vsy} adapted to the $N_c=2$ case.

Using the properties of the Bessel function we find the following asymptotic behaviors:
\begin{eqnarray}
\frac{\lambda_Q^c(\mathbf{p})}{\lambda_{Q_T}^c(\mathbf{p})}&\to & 1\ \ \mathrm{for}\ \ Q\to Q_T\ \ \mathrm{or}\ \ p\to 0, \\
\frac{\lambda_Q^c(\mathbf{p})}{\lambda_{Q_T}^c(\mathbf{p})}&\to & \left(\frac{Q_T}{Q}\right)^\frac{\alpha_s \beta_0^g}{2 \pi}\ \ \mathrm{for}\ \ Q\to 0\ \ \mathrm{or}\ \ p\to \infty.
\end{eqnarray}
{\color{black} The asymptotic behavior for $Q\rightarrow Q_T$ is easy to understand as in this case there is no evolution and $\lambda$ remains unchanged. For $\mathbf{p}\rightarrow 0$ the reason for the same asymptotics is that a target field with very low transverse momentum does not resolve the components of the dressed gluon state, which therefore scatters like a pointlike bare gluon. For large $\mathbf{p}$ the resolution is maximal -- the substructure is fully resolved and $\lambda$ attains its maximal value at a given $Q$, which must grow as $Q$ is taken smaller.}

\subsection{Solving for $\Delta$}
The equation for $\Delta$ reads
\begin{multline}
    \frac{\partial}{\partial\ln Q^2}\Delta_Q(\mathbf{z})=-\frac{\alpha_s\beta^g_0}{4\pi}\left[1+\Delta_Q(\mathbf{z})+\frac{1}{2}\int{\frac{d\phi}{2\pi}\left(-2-2\Delta_Q(\mathbf{z}+Q^{-1}\mathbf{e}_r/2)\right.}\right.\\\left.{\left.-2\Delta_Q(\mathbf{z}-Q^{-1}\mathbf{e}_r/2)-\frac{2}{3}\lambda_Q^c(\mathbf{z}+Q^{-1}\mathbf{e}_r/2)\lambda_Q^c(\mathbf{z}-Q^{-1}\mathbf{e}_r/2)\right)}\right].
\end{multline}
We again Fourier transform it to momentum space.
For the last term we get
\begin{eqnarray}
    -\frac{2}{3}\int{\frac{d\phi}{2\pi}\lambda_Q^c(\mathbf{z}_1)\lambda_Q^c(\mathbf{z}_2)}&=&-\frac{2}{3}\int{\frac{d\phi}{2\pi}\int{\frac{d^2\mathbf{k}_1d^2\mathbf{k}_2}{(2\pi)^4}e^{i(\mathbf{k}_1+\mathbf{k}_2)\mathbf{z}}e^{i\frac{1}{2Q}(\mathbf{k}_1-\mathbf{k}_2)\cdot\mathbf{e}_r}\lambda_Q^c(\mathbf{k}_1)\lambda_Q^c(\mathbf{k}_2)}}\nonumber\\
    &=&-\frac{2}{3}\int{\frac{d\phi}{2\pi}\int{\frac{d^2\mathbf{p}d^2\mathbf{k}}{(2\pi)^4}e^{i\mathbf{p}\mathbf{z}}e^{i\frac{1}{Q}\mathbf{k}\cdot\mathbf{e}_r}\lambda_Q^c(\mathbf{p}/2+\mathbf{k})\lambda_Q^c(\mathbf{p}/2-\mathbf{k})}}\nonumber\\
    &=&-\frac{2}{3}\int{\frac{d^2\mathbf{p}d^2\mathbf{k}}{(2\pi)^4}e^{i\mathbf{p}\mathbf{z}}J_0(Q^{-1}\lvert\mathbf{k}\rvert)\lambda_Q^c(\mathbf{p}/2+\mathbf{k})\lambda_Q^c(\mathbf{p}/2-\mathbf{k})}.
\end{eqnarray}
The equation then becomes
\begin{eqnarray}
    \frac{\partial}{\partial\ln Q^2}\Delta_Q(\mathbf{p})&=&-\frac{\alpha_s\beta^g_0}{4\pi}\left[\delta^{(2)}(\mathbf{p})+\Delta_Q(\mathbf{p})+\frac{1}{2}\left(-2\delta^{(2)}(\mathbf{p})-4J_0\left(\frac{p}{2Q}\right)\Delta_Q(\mathbf{p})\right.\right.\nonumber\\
    &&\hskip 2cm\left.\left.-\frac{2}{3}\int{\frac{d^2\mathbf{k}}{(2\pi)^2}J_0(Q^{-1}\lvert\mathbf{k}\rvert)\lambda_Q^c(\mathbf{p}/2+\mathbf{k})\lambda_Q^c(\mathbf{p}/2-\mathbf{k})}\right)\right]\nonumber \\
&=&-\frac{\alpha_s\beta^g_0}{4\pi}\left[\Delta_Q(\mathbf{p})-2J_0\left(\frac{p}{2Q}\right)\Delta_Q(\mathbf{p})\right.\nonumber\\
&&\hskip 2cm\left.-\frac{1}{3}\int{\frac{d^2\mathbf{k}}{(2\pi)^2}J_0(Q^{-1}k)\lambda_Q^c(\mathbf{p}/2+\mathbf{k})\lambda_Q^c(\mathbf{p}/2-\mathbf{k})}\right],
\end{eqnarray}
with $\lvert\mathbf{k}\rvert=k$.
Defining the source term
\begin{equation}\label{F}
    F(\mathbf{p},Q)=\frac{\alpha_s\beta^g_0}{12\pi}\int{\frac{d^2\mathbf{k}}{(2\pi)^2}J_0(Q^{-1}k)\lambda_Q^c(\mathbf{p}/2+\mathbf{k})\lambda_Q^c(\mathbf{p}/2-\mathbf{k})},
\end{equation}
we get
\begin{equation}
    \frac{\partial}{\partial\ln Q^2}\Delta_Q(\mathbf{p})=-\frac{\alpha_s\beta^g_0}{4\pi}\left[\Delta_Q(\mathbf{p})-2J_0\left(\frac{p}{2Q}\right)\Delta_Q(\mathbf{p})\right]+F(\mathbf{p},Q).
\end{equation}

The homogeneous solution to this equation reads 
\begin{equation}
\Delta_Q^h(\mathbf{p})=\exp\left[-\int_Q^{Q_T}{\frac{dQ^2}{Q^2}R_\Delta(p,Q)}\right]\Delta_{Q_T}^h(\mathbf{p}),
\end{equation}
where
\begin{equation}
R_\Delta(p,Q)=\frac{\alpha_s\beta^g_0}{4\pi}\left[2J_0\left(\frac{p}{2Q}\right)-1\right].
\end{equation}

Including the source term yields the solution of the equation with the appropriate initial condition as
\begin{eqnarray}
\label{eq:solforDelta}
\Delta_Q(\mathbf{p})&=&\exp\left[-\int_Q^{Q_T}{\frac{dQ'^2}{Q'^2}R_\Delta(p,Q')}\right]\Delta_{Q_T}(\mathbf{p})\\&&-\int_Q^{Q_T}{\frac{dQ'^2}{Q'^2}\exp\left[-\int_{Q}^{Q'}{\frac{dQ''^2}{Q''^2}R_\Delta(p,Q'')}\right]F(\mathbf{p},Q')},\nonumber
\end{eqnarray}
with the source function $F$ given by \eqref{F}.

Asymptotic estimates of the behavior of some quantities are:
\begin{eqnarray}
\exp\left[-\int_Q^{Q_T}{\frac{dQ'^2}{Q'^2}R_\Delta(p,Q')}\right]&\to & 1\ \ \mathrm{for}\ \ Q\to Q_T, \\
\exp\left[-\int_Q^{Q_T}{\frac{dQ'^2}{Q'^2}R_\Delta(p,Q')}\right]&\to & \left(\frac{Q_T}{Q}\right)^\frac{\alpha_s \beta_0^g}{2 \pi}\ \ \mathrm{for}\ \ Q\to 0\ \ \mathrm{and/or}\ \ p\to \infty,\\
\exp\left[-\int_Q^{Q_T}{\frac{dQ'^2}{Q'^2}R_\Delta(p,Q')}\right]&\to & \left(\frac{Q}{Q_T}\right)^\frac{\alpha_s \beta_0^g}{2 \pi}\ \ \mathrm{for}\ \ p\to 0;
\end{eqnarray}
For the function
\begin{equation}
\mathcal{A}(p,Q)=\int_Q^{Q_T}{\frac{dQ'^2}{Q'^2}\exp\left[-\int_{Q}^{Q'}{\frac{dQ''^2}{Q''^2}R_\Delta(p,Q'')}\right]F(\mathbf{p},Q')},
\end{equation}
we find
\begin{eqnarray}
\mathcal{A}(p,Q)&\to & 0\ \ \mathrm{for}\ \ Q\to Q_T\ \ \mathrm{or}\ \ Q\to 0, \\
\mathcal{A}(p,Q)&\to & 0\ \mathrm{for}\ p\to \infty,\ \ Q=Q_T.
\end{eqnarray}
We also find that for $p\rightarrow 0$, function $\mathcal{A}(p,Q)$ is a monotonically increasing function of $Q$.
\subsection{Solving for $B$}
The equation for $B$ reads
\begin{multline}
    \frac{\partial}{\partial\ln Q^2}B_Q^{cd}(\mathbf{z})=\frac{\alpha_s\beta^g_0}{4\pi}\left[B_Q^{cd}(\mathbf{z})+\frac{1}{2}\int\frac{d\phi}{2\pi}\left(B_Q^{cd}(\mathbf{z}_2)+B_Q^{cd}(\mathbf{z}_1)+\lambda_Q^c(\mathbf{z}_1)\lambda_Q^d(\mathbf{z}_2)\right.\right.\\\left.\left.-\frac{1}{3}\delta_{cd}\lambda_Q^a(\mathbf{z}_1)\lambda_Q^a(\mathbf{z}_2)\right)\right],
\end{multline}
which in momentum space becomes
\begin{multline}
    \frac{\partial}{\partial\ln Q^2}B_Q^{cd}(\mathbf{p})=-\frac{\alpha_s\beta^g_0}{4\pi}\left[B_Q^{cd}(\mathbf{p})+\frac{1}{2}\left(2J_0\left(\frac{p}{2Q}\right)B_Q^{cd}(\mathbf{p})\right.\right.\\\left.\left.+\int{\frac{d^2\mathbf{k}}{(2\pi)^2}J_0(Q^{-1}k)\left[\lambda_Q^c(\mathbf{p}/2+\mathbf{k})\lambda_Q^d(\mathbf{p}/2-\mathbf{k})-\frac{1}{3}\delta_{cd}\lambda_Q^a(\mathbf{p}/2+\mathbf{k})\lambda_Q^a(\mathbf{p}/2-\mathbf{k})\right]}\right)\right].
\end{multline}
Defining the source
\begin{eqnarray}\label{G}
    G^{cd}(\mathbf{p},Q)&=&-\frac{\alpha_s\beta^g_0}{4\pi N_c}\int\frac{d^2\mathbf{k}}{(2\pi)^2}J_0(Q^{-1}k)\left[\lambda_Q^c(\mathbf{p}/2+\mathbf{k})\lambda_Q^d(\mathbf{p}/2-\mathbf{k})\right.\nonumber\\&&\hskip 2cm \left.-\frac{1}{3}\delta_{cd}\lambda_Q^a(\mathbf{p}/2+\mathbf{k})\lambda_Q^a(\mathbf{p}/2-\mathbf{k})\right],
\end{eqnarray}
we arrive at
\begin{equation}
    \frac{\partial}{\partial\ln Q^2}B_Q^{cd}(\mathbf{p})=-\frac{\alpha_s\beta^g_0}{4\pi}\left[B_Q^{cd}(\mathbf{p})+J_0\left(\frac{p}{2Q}\right)B_Q^{cd}(\mathbf{p})\right]+G^{cd}(\mathbf{p},Q).
\end{equation}
The solution is analogous to~\eqref{eq:solforDelta}:
\begin{eqnarray}
\label{eq:solforB}
    B_Q^{cd}(\mathbf{p})&=&\exp\left[-\int_Q^{Q_T}{\frac{dQ'^2}{Q'^2}R_B(p,Q')}\right]B^{cd}_{Q_T}(\mathbf{p})\nonumber \\&&-\int_Q^{Q_T}{\frac{dQ'^2}{Q'^2}\exp\left[-\int_{Q}^{Q'}{\frac{dQ''^2}{Q''^2}R_B(p,Q'')}\right]G^{cd}(\mathbf{p},Q')},
\end{eqnarray}
with
\begin{equation}
    R_B(p,Q)=-\frac{\alpha_s\beta^g_0}{4\pi}\left[J_0\left(\frac{p}{2Q}\right)+1\right]
\end{equation}
and $G$ given by \eqref{G}.

Asymptotic estimates of the behavior of some quantities are:
\begin{eqnarray}
\exp\left[-\int_Q^{Q_T}{\frac{dQ'^2}{Q'^2}R_B(p,Q')}\right]&\to & 1\ \ \mathrm{for}\ \ Q\to Q_T, \\
\exp\left[-\int_Q^{Q_T}{\frac{dQ'^2}{Q'^2}R_B(p,Q')}\right]&\to & \left(\frac{Q_T}{Q}\right)^\frac{\alpha_s \beta_0^g}{2 \pi}\ \ \mathrm{for}\ \ Q\to 0\ \ \mathrm{and/or}\ \ p\to \infty,\\
\exp\left[-\int_Q^{Q_T}{\frac{dQ'^2}{Q'^2}R_B(p,Q')}\right]&\to & \left(\frac{Q_T}{Q}\right)^\frac{\alpha_s \beta_0^g}{\pi}\ \ \mathrm{for}\ \ p\to 0;
\end{eqnarray}
For the function
\begin{equation}
\mathcal{B}^{cd}(p,Q)=\int_Q^{Q_T}{\frac{dQ'^2}{Q'^2}\exp\left[-\int_{Q}^{Q'}{\frac{dQ''^2}{Q''^2}R_B(p,Q'')}\right]G^{cd}(\mathbf{p},Q')},
\end{equation}
we find
\begin{eqnarray}
\mathcal{B}^{cd}(p,Q)&\to & 0\ \ \mathrm{for}\ \ Q\to Q_T\ \ \mathrm{or}\ \ Q\to 0, \\
\mathcal{B}^{cd}(p,Q)&\to & 0 \ \mathrm{for}\ p\to \infty,\ \ Q=Q_T.
\end{eqnarray}
Besides, function $\mathcal{B}^{cd}(p,Q)$ at $p\rightarrow 0$ is a monotonically decreasing function of $Q$.

\section{Initial condition: A single dipole target}

To understand the behavior of the solution obtained above, we have to choose a physical initial condition. {\color{black} As discussed in the introduction, the situation in which the dilute approximation is appropriate is typical of dipole-dipole scattering. We thus} choose the initial condition that corresponds to a single dipole target. {\color{black} In the dilute case the eikonal $S$ matrix of a gluon in the light cone gauge is given by
\begin{equation}\label{sl}
    S(\mathbf {z})=1+igT^aa^{a}(\mathbf{z}),
\end{equation}
where $a^a(\mathbf{z})$ is the color scalar potential of the target. 
}The color field of a single dipole at points $\mathbf{x}_1$ and $\mathbf{x}_2$ is just a superposition of the fields of these two charges. {\color{black} Choosing a fundamental dipole oriented in the third direction in color space, the color field of the dipole is
\begin{equation}
a^a(\mathbf{z})=\frac{g}{2\pi^{3/2}}\delta^{a3}\left[\ln{(\mathbf{x}_{1}-\mathbf{z})^2/L^2}
-\ln{(\mathbf{x}_{2}-\mathbf{z})^2/L^2}\right],
\end{equation}
where $L$ is an IR cutoff necessary to regulate the potential of an individual color charge, but which cancels for a dipole.

The $S$-matrix for scattering on such a field is determined by the contribution due to $\lambda$ via \eqref{sl}. Taking Fourier transform with respect to $\mathbf{z}$ we find}
$\lambda_{Q_T}^a(\mathbf{p})\propto\delta^{3a}\frac{1}{p^2}(e^{i\mathbf{p}\mathbf{x}_1}-e^{i\mathbf{p}\mathbf{x}_2})$. Choosing $\mathbf{x}_1=-\mathbf{x}_2=\mathbf{x}$ we have
\begin{equation}\label{incl}\lambda_{Q_T}^a(\mathbf{p})=\lambda\delta^{3a}\frac{1}{p^2}\sin(\mathbf{p}\cdot \mathbf{x}).
\end{equation}
{\color{black} The constant $\lambda$ here is of the order of $\alpha_s$. Its exact value depends on the representation of the target dipole, but it is unimportant in our calculations as it only determines the overall scale of the field and cancels in the ratios that we calculate in the following.}

At the initial scale $Q_T$ we write
\begin{equation}
    \mathbb{S}_{Q_T}^{ab}(\mathbf{z})=(1+\Delta_{Q_T}(\mathbf{z}))\delta_{ab}-\lambda_{Q_T}^c(\mathbf{z})\epsilon_{abc}-B_{Q_T}^{ab}(\mathbf{z})\ ,
\end{equation}
For small $\lambda$ as before
$\Delta_{Q_T}(\mathbf{z}), B_{Q_T}^{cd}\propto\lambda^2$. 

The matrix $\mathbb{S}_{Q_T}^{ab}(\mathbf{p})$ should be unitary. To order $\lambda^2$ the unitarity condition ($\mathbb{S}_{Q_T}^{ab}(\mathbf{z})\mathbb{S}_{Q_T}^{bc}{}^\dagger(\mathbf{z})=\delta^{ac}+\mathcal{O}(\lambda^3)$) becomes
\begin{eqnarray}\Delta_{Q_T}(\mathbf{z})&=&-\frac{1}{2}\lambda_{Q_T}^c(\mathbf{z})\lambda_{Q_T}^c(\mathbf{z}),\\
B_{Q_T}^{ab}(\mathbf{z})&=&-\frac{1}{4}\left(\lambda_{Q_T}^a(\mathbf{z})\lambda_{Q_T}^b(\mathbf{z})-\frac{1}{3}\delta^{ab}\lambda_{Q_T}^c(\mathbf{z})\lambda_{Q_T}^c(\mathbf{z})\right).
\end{eqnarray}
In momentum space 
\begin{eqnarray}
\Delta_{Q_T}(\mathbf{p})&=&-\frac{1}{2}\int{\frac{d^2\mathbf{k}}{(2\pi)^2}\lambda_{Q_T}^c(\mathbf{k})\lambda_{Q_T}^c(\mathbf{p}-\mathbf{k})} ,   \\
B_{Q_T}^{ab}(\mathbf{p})&=&-\frac{1}{4}\left(\int{\frac{d^2\mathbf{k}}{(2\pi)^2}\lambda_{Q_T}^a(\mathbf{k})\lambda_{Q_T}^b(\mathbf{p}-\mathbf{k})}-\frac{1}{3}\delta^{ab}\int{\frac{d^2\mathbf{k}}{(2\pi)^2}\lambda_{Q_T}^c(\mathbf{k})\lambda_{Q_T}^c(\mathbf{p}-\mathbf{k})}\right).
\end{eqnarray}

It is convenient to introduce
\begin{equation}
\Delta_{Q}^U(\mathbf{p})=-\frac{1}{2}\int{\frac{d^2\mathbf{k}}{(2\pi)^2}\lambda_{Q}^c(\mathbf{k})\lambda_{Q}^c(\mathbf{p}-\mathbf{k})}\ .
\end{equation}
In terms of this quantity the deviation of the evolved $S$-matrix from unitarity can be estimated by considering
$1-\frac{\Delta_{Q}^U(\mathbf{p})}{\Delta_{Q}(\mathbf{p})}$.

Since for our initial condition $\lambda_Q^{c}\propto\delta^{c3}$ holds at any $Q$, the matrix $B_Q^{ab}(\mathbf{p})$ is diagonal and can be written as 
\begin{equation}
    B_{Q}^{ab}(\mathbf{p})=B_Q(\mathbf{p})\left(\delta^{a3}\delta^{b3}-\frac{1}{3}\delta^{ab}\right),
\end{equation}
with 
\begin{eqnarray}
    B_Q(\mathbf{p})&=&\exp\left[-\int_Q^{Q_T}{\frac{dQ'^2}{Q'^2}R_B(p,Q')}\right]B_{Q_T}(\mathbf{p})\nonumber \\&&-\int_Q^{Q_T}{\frac{dQ'^2}{Q'^2}\exp\left[-\int_{Q}^{Q'}{\frac{dQ''^2}{Q''^2}R_B(p,Q'')}\right]G(\mathbf{p},Q')}.
\end{eqnarray}
Here
\begin{equation}
    G(\mathbf{p},Q)=-\frac{\alpha_s\beta^g_0}{8\pi}\int{\frac{d^2\mathbf{k}}{(2\pi)^2}J_0(Q^{-1}k)\left[\lambda_Q^a(\mathbf{p}/2+\mathbf{k})\lambda_Q^a(\mathbf{p}/2-\mathbf{k})\right]}
\end{equation}
and
\begin{equation}
    B_{Q_T}(\mathbf{p})=-\frac{1}{4}\int{\frac{d^2\mathbf{k}}{(2\pi)^2}\lambda_{Q_T}^a(\mathbf{k})\lambda_{Q_T}^a(\mathbf{p}-\mathbf{k})}.
    \end{equation}

Introducing
\begin{equation}
    B_{Q}^U(\mathbf{p})=-\frac{1}{4}\int{\frac{d^2\mathbf{k}}{(2\pi)^2}\lambda_{Q}^a(\mathbf{k})\lambda_{Q}^a(\mathbf{p}-\mathbf{k})},
    \end{equation}
another indicator of the deviation from unitarity is 
$1-\frac{B_{Q}^U(\mathbf{p})}{B_{Q}(\mathbf{p})}$.

\subsection{Deviation from unitarity}
The convenient dimensionless measures of the amount of the evolution are given by 
\begin{eqnarray}
    \frac{\Delta_{Q_T}(\mathbf{p})-\Delta_{Q}(\mathbf{p})}{\Delta_{Q_T}(\mathbf{p})}&=&1-\frac{\Delta_{Q}(\mathbf{p})}{\Delta_{Q_T}(\mathbf{p})}\ ,\\
\frac{B_{Q_T}(\mathbf{p})-B_{Q}(\mathbf{p})}{B_{Q_T}(\mathbf{p})}&=&1-\frac{B_{Q}(\mathbf{p})}{B_{Q_T}(\mathbf{p})}\ .
\end{eqnarray}

To quantify the deviation of the matrix $\mathbb{S}_Q$ from unitarity during the evolution, we consider the following quantities:
\begin{eqnarray}
\label{eq:rdelta}    \mathcal{R}_\Delta(Q,\mathbf{p})&=&\frac{1-\Delta^U_Q(\mathbf{p})/\Delta_{Q}(\mathbf{p})}{1-\Delta_Q(\mathbf{p})/\Delta_{Q_T}(\mathbf{p})}\ ,\label{Rd}\\
\label{eq:rb} 
\mathcal{R}_B(Q,\mathbf{p})&=&\frac{1-B^U_Q(\mathbf{p})/B_{Q}(\mathbf{p})}{1-B_Q(\mathbf{p})/B_{Q_T}(\mathbf{p})}\ .
\end{eqnarray}
 If the deviation from unitarity is small ($\Delta_Q\approx \Delta^U_Q$, and the same for $B$) while the evolution is sizeable ($\Delta_Q\ne \Delta_{Q_T}$, and the same for $B$), these quantities are close to zero. If, on the other hand the deviation from unitarity is of the same order as the amount of evolution, the absolute value of each ratio should be close to unity. 

\section{Numerical results}

\subsection{The evolution}
\begin{figure}[hbt]
\centering
\includegraphics[width=0.5\textwidth]{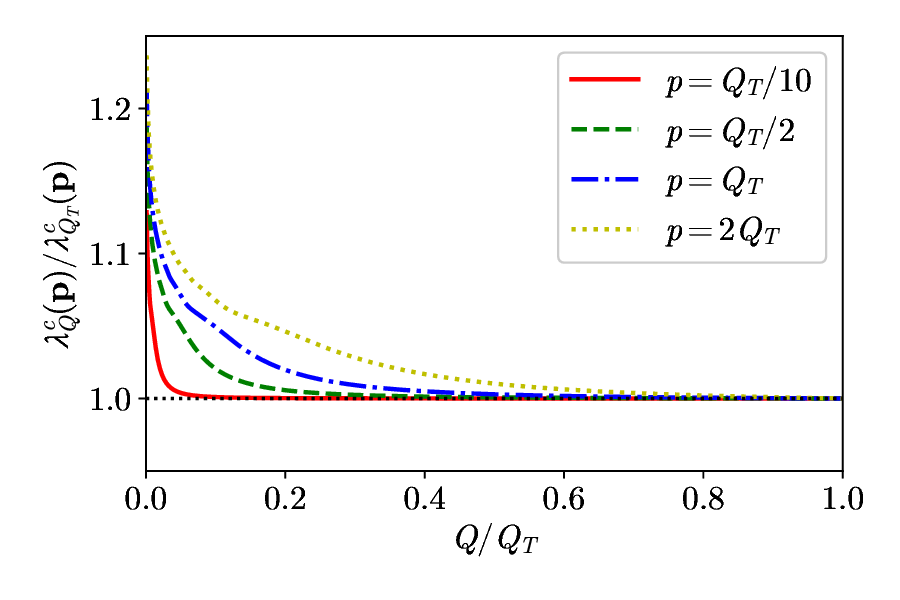}\includegraphics[width=0.5\textwidth]{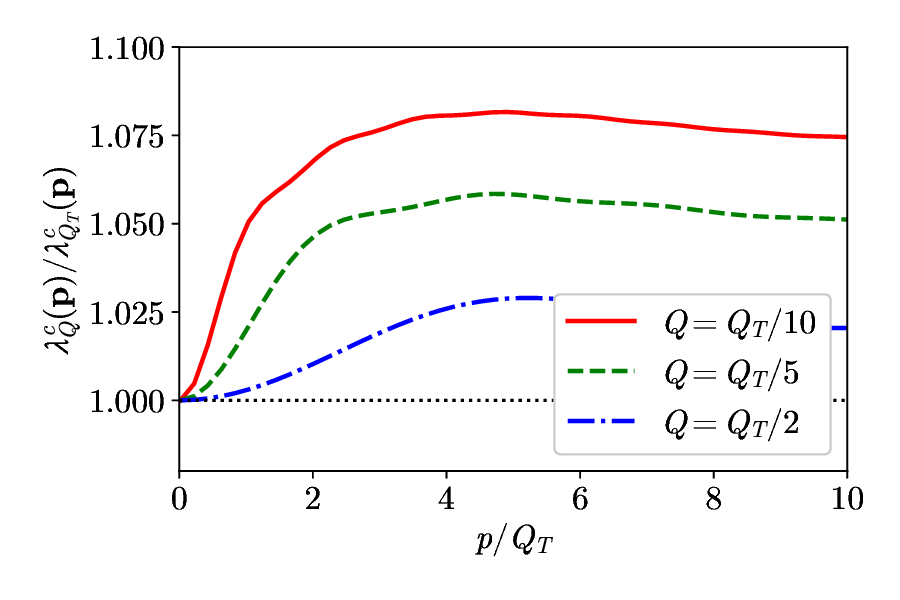}
\caption{Evolution of $\lambda_Q(\bf{p})$ with $Q$. $\lambda^c_Q/\lambda^c_{Q_T}$ versus $Q/Q_T$ for different values of $p/Q_T$ (left), and versus $p/Q_T$ for different values of $Q/Q_T$ (right) at $\alpha_s=0.1$. The angle of the momentum $\bf{p}$ relative to the direction of the dipole is taken to be $\cos\phi=0$.}
\label{fig:fig1}
\end{figure}

We have calculated the solutions  $\lambda_Q$, $\Delta_Q$ and $B_Q$~\eqref{eq:sollambda},~\eqref{eq:solforDelta},~\eqref{eq:solforB} numerically for the initial condition~\eqref{incl} and have studied deviations from unitarity for various values of the QCD coupling $\alpha_s$. On Fig.~\ref{fig:fig1} we plot the evolution of $\lambda_Q$. As expected, we observe that the deviation from the initial condition is most pronounced at small values of $Q$. For $\alpha_s=0.1$ this deviation is rather modest and reaches at most $5-8\%$ down to $Q=Q_T/10$.

\begin{figure}[hbt]
\centering
\includegraphics[width=0.5\textwidth]{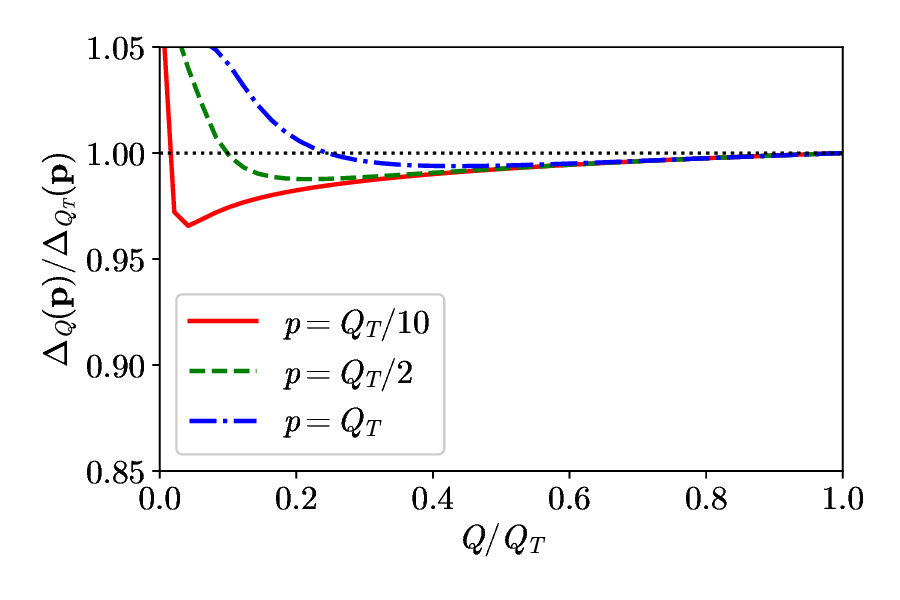}\includegraphics[width=0.5\textwidth]{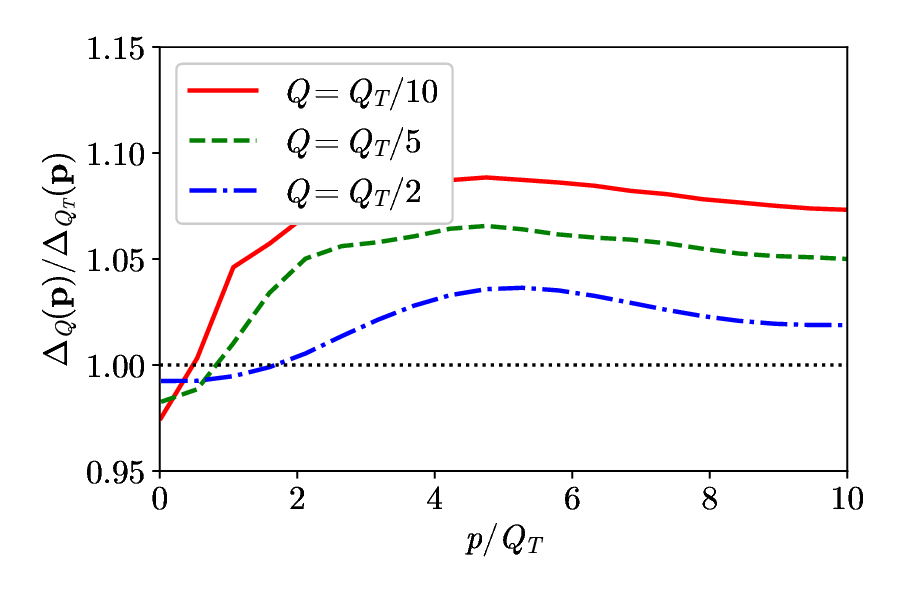}
\caption{$\Delta_Q/\Delta_{Q_T}$ versus $Q/Q_T$ for different values of $p/Q_T$ (left), and versus $p/Q_T$ for different values of $Q/Q_T$ (right), for $\alpha_s=0.1$ and $\cos\phi=0$.}
\label{fig:fig2}
\end{figure}

On Fig.~\ref{fig:fig2}  we plot the same for $\Delta_Q$. The picture is qualitatively similar. Deviations from initial conditions are under 10\% unless one goes to extremely small values of $Q$.

\begin{figure}[hbt]
\centering
\includegraphics[width=0.8\textwidth]{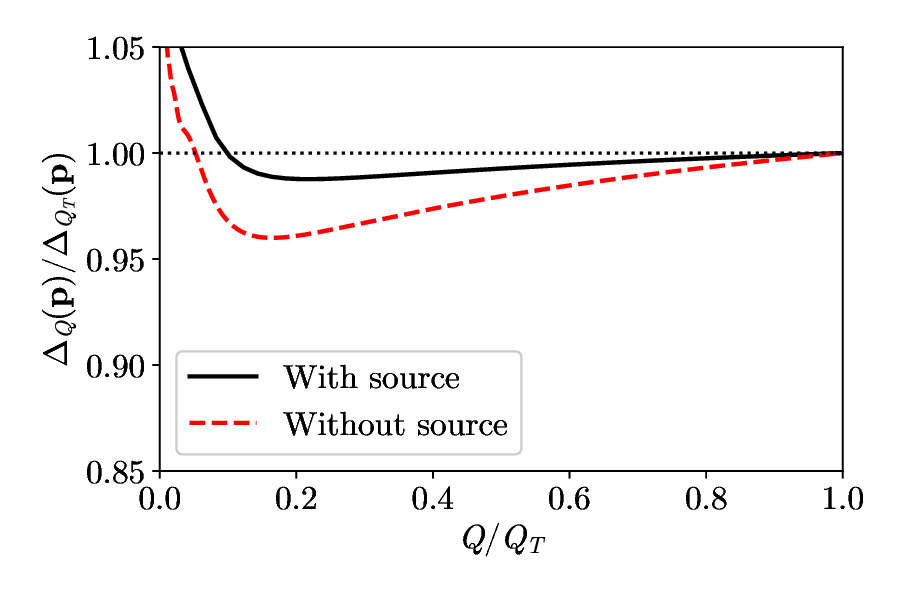}
\caption{$\Delta_Q/\Delta_{Q_T}$ versus $Q/Q_T$ with and without the source term $F$ in~\eqref{eq:solforDelta}, for $\alpha_s=0.1$ and $\cos\phi=0$.}
\label{fig:fig3}
\end{figure}

Fig.~\ref{fig:fig3} illustrates the importance of the source term $F$ due to $\lambda$ in the evolution of $\Delta$, eq.~\eqref{eq:solforDelta}. Interestingly, the presence of $F$ significantly tames the growth of $\Delta$ away from the initial condition. 

\begin{figure}[hbt]
\centering
\includegraphics[width=0.5\textwidth]{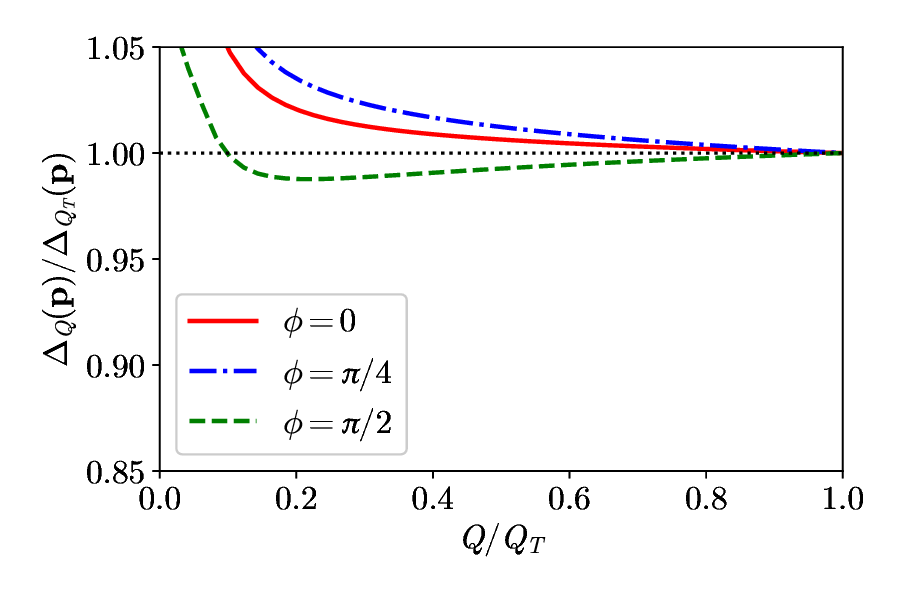}\includegraphics[width=0.5\textwidth]{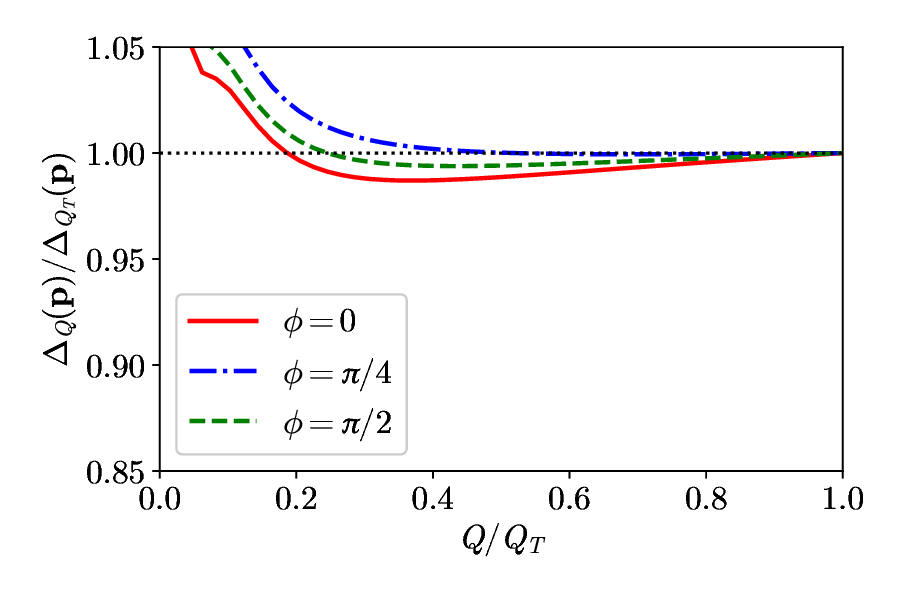}
\caption{$\Delta_Q/\Delta_{Q_T}$ versus $Q/Q_T$ for different  dipole orientations, for $p=Q_T/2$ (left) and $p=Q_T$ (right), for $\alpha_s=0.1$.}
\label{fig:fig4}
\end{figure}

Fig.~\ref{fig:fig4} shows the dependence on the orientation between the momentum and the original dipole in the initial condition~\eqref{incl}. The dependence on the orientation is nontrivial, e.g., the sign of the evolved $\Delta$ can be negative or  positive, depending on the angle. The  qualitative picture however remains the same -- the effect of the evolution is perhaps surprisingly small even for $Q\sim Q_T/5$.

\begin{figure}[hbt]
\centering
\includegraphics[width=0.5\textwidth]{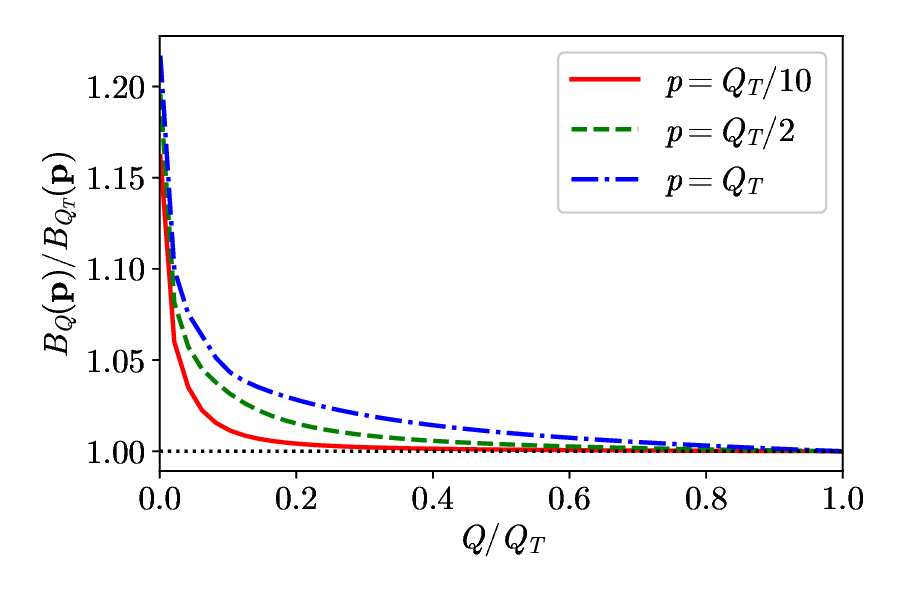}\includegraphics[width=0.5\textwidth]{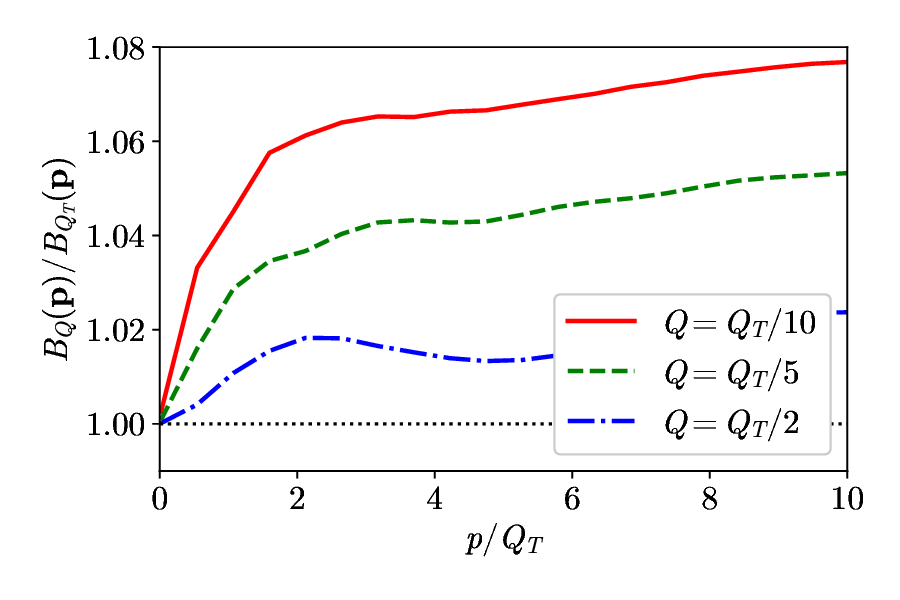}
\caption{$B_Q/B_{Q_T}$ versus $Q/Q_T$ for different values of $p/Q_T$ (left), and versus $p/Q_T$ for different values of $Q/Q_T$ (right), for $\alpha_s=0.1$ and $\cos\phi=0$.}
\label{fig:fig5}
\end{figure}

Finally, in Fig.~\ref{fig:fig5} we plot the dependence on $Q$ and $p$ for $B_Q$. The picture is qualitatively similar to that for $\Delta_Q$.

\subsection{Deviations from unitarity}
\begin{figure}[htp]
\centering
\includegraphics[width=0.5\textwidth]{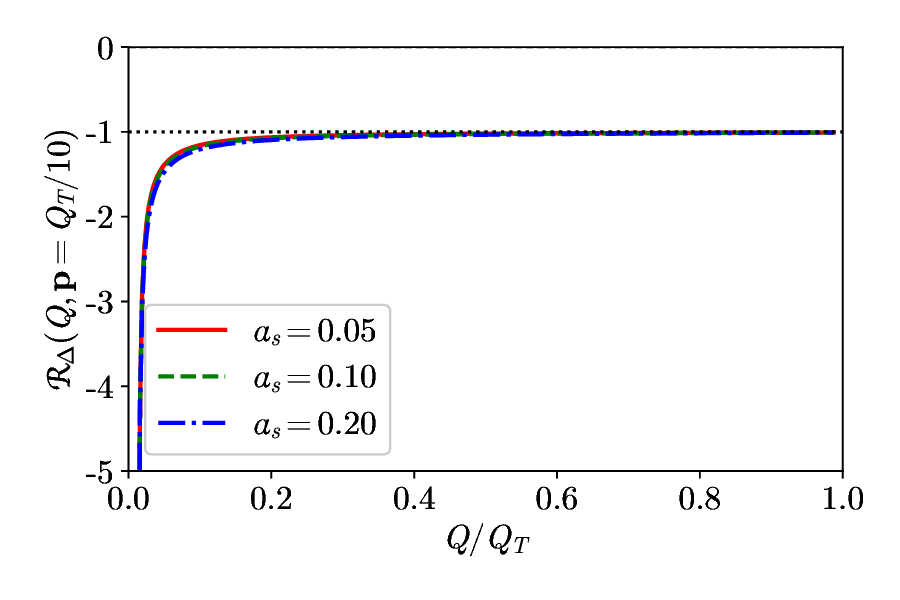}\includegraphics[width=0.5\textwidth]{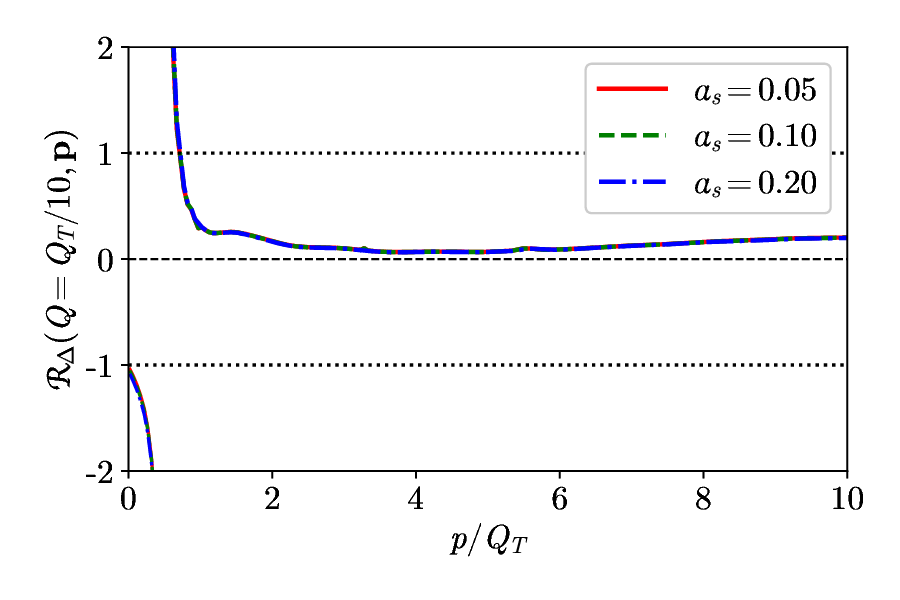}
\includegraphics[width=0.5\textwidth]{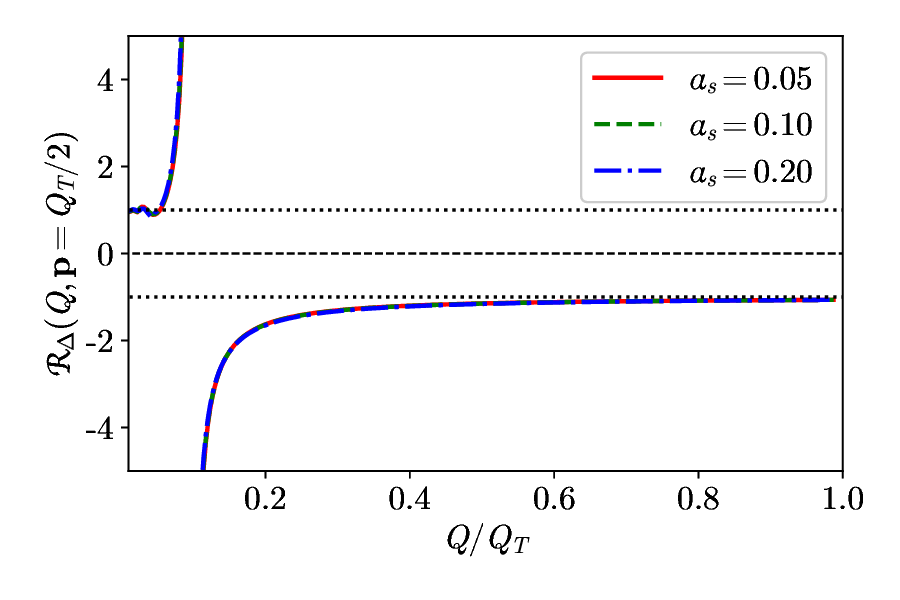}\includegraphics[width=0.5\textwidth]{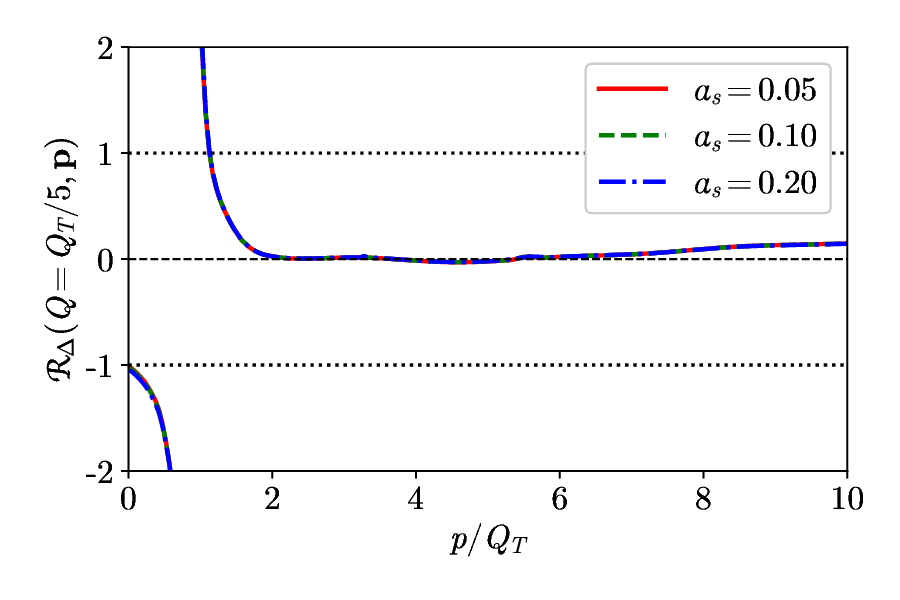}
\includegraphics[width=0.5\textwidth]{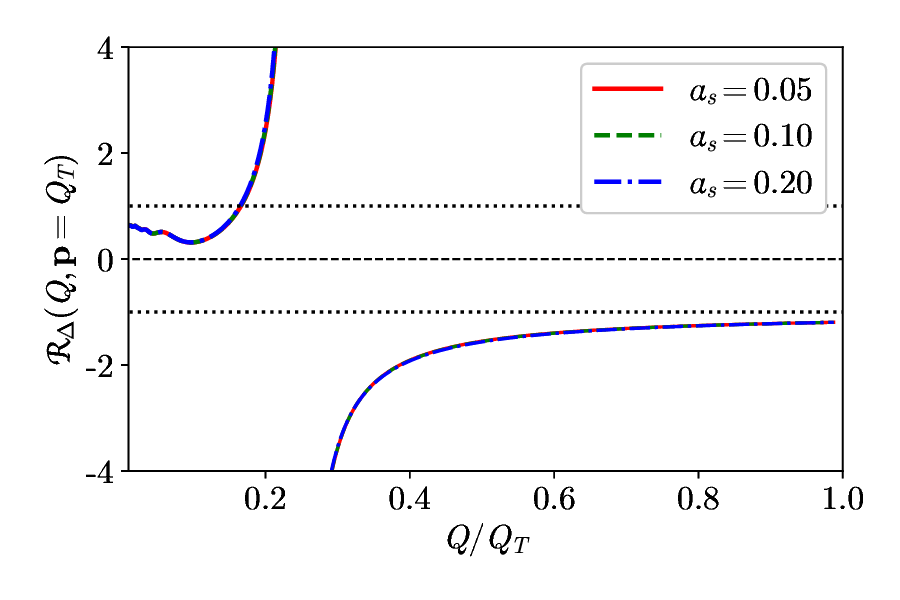}\includegraphics[width=0.5\textwidth]{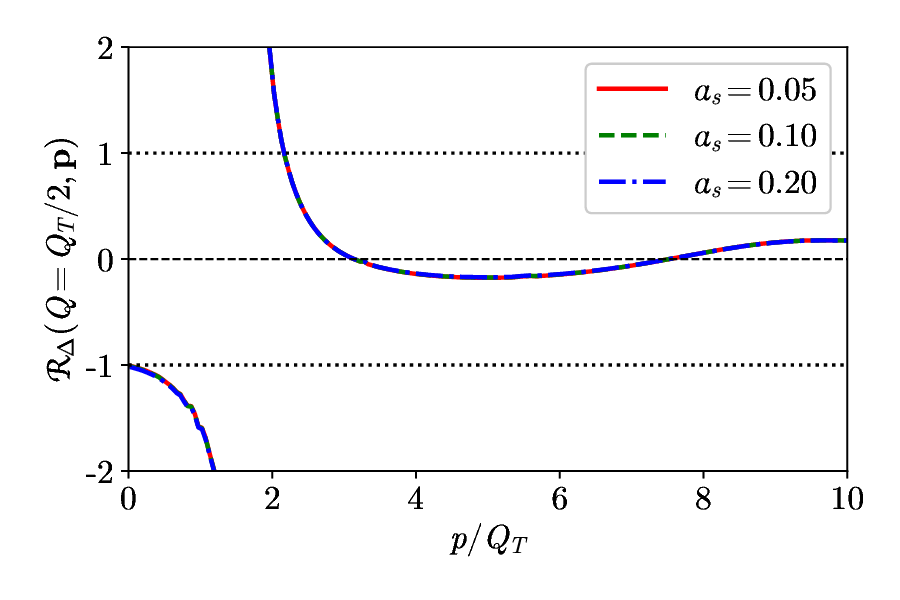}
\caption{$\mathcal{R}_\Delta(Q,\mathbf{p})$ for $p=Q_T/10$ (top), $p=Q_T/2$ (middle) and $p=Q_T$ (bottom) versus $Q/Q_T$ (left plots), and for $Q=Q_T/10$ (top), $Q=Q_T/5$ (middle) and $Q=Q_T/2$ (bottom) versus $p/Q_T$ (right plots), for different values of $\alpha_s$, for $\cos\phi=0$.}
    \label{fig:RDeltaQ}
\end{figure}

We now study the deviation of $\mathbb{S}_Q$ from unitarity. On Fig.~\ref{fig:RDeltaQ} we plot the ratio $R_\Delta$ for different values of $p$ as a function of $Q$ (left panel) and for different values of $Q$ as a function of $p$ (right panel).

We observe that the deviations from unitarity are generically of order one for values of momenta $p< Q_T$ for $Q$ that is not too small, i.e., $Q> Q_T/3$. In fact in this range of momenta the ration $R_\Delta$ is very close to negative unity, indicating that the deviation from unitarity is equal in magnitude and opposite in sign to the amount of the evolution away from the initial condition. On the other hand for larger values of $p$ the ratio is close to zero and the matrix $\mathbb{S}_Q$ is close to unitary. The discontinuity observed in most plots on Fig.~\ref{fig:RDeltaQ} is clearly due to "accidental" vanishing of the amount of evolution at some value of $p$ for (almost) every $Q$ (and conversely for some value of $Q$ for every $p$).

A stand out property of these curves is their practical independence on the value of the coupling constant. Varying the value of coupling by a factor of four, practically does not affect the curves. We will discuss this "scaling" property a little later.

\begin{figure}[htp]
\centering
\includegraphics[width=0.5\textwidth]{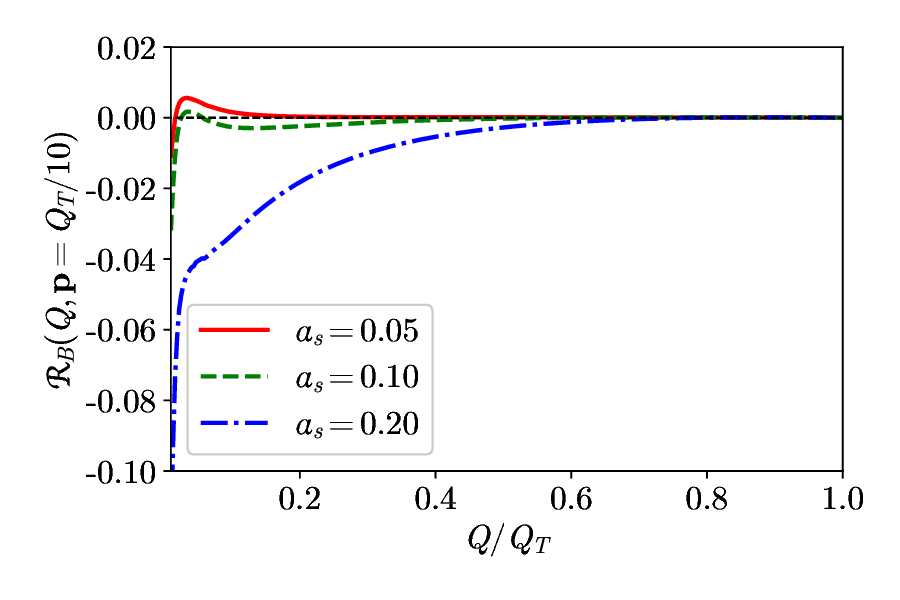}\includegraphics[width=0.5\textwidth]{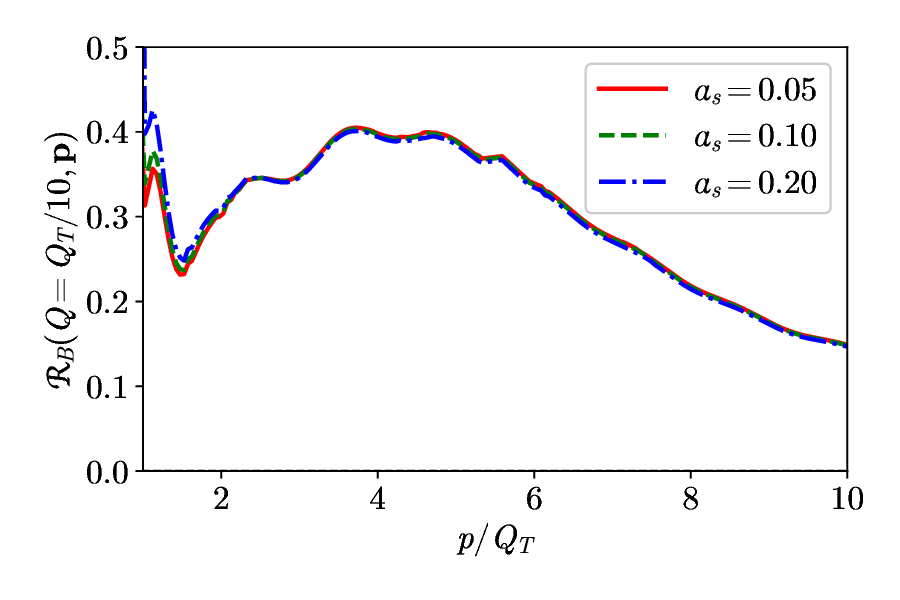}
\includegraphics[width=0.5\textwidth]{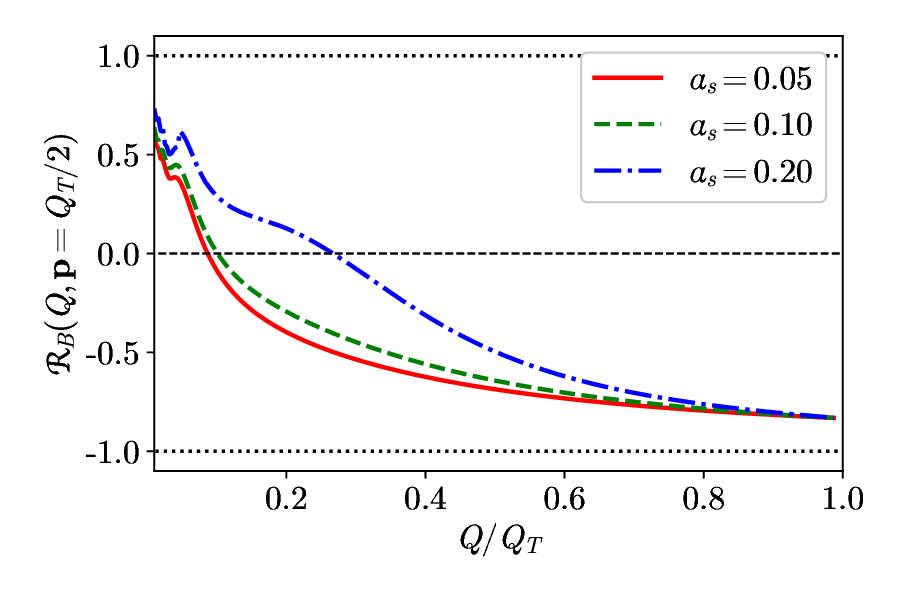}\includegraphics[width=0.5\textwidth]{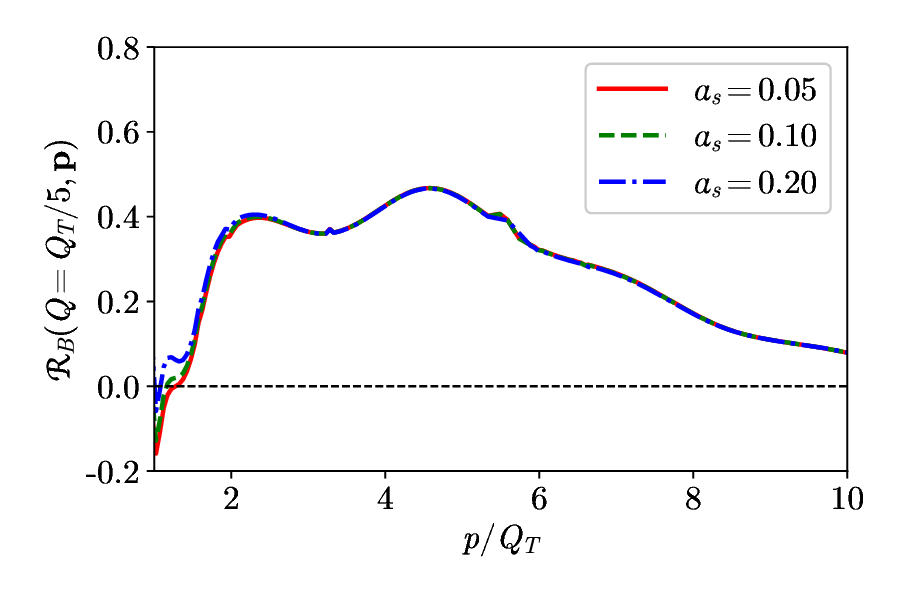}
\includegraphics[width=0.5\textwidth]{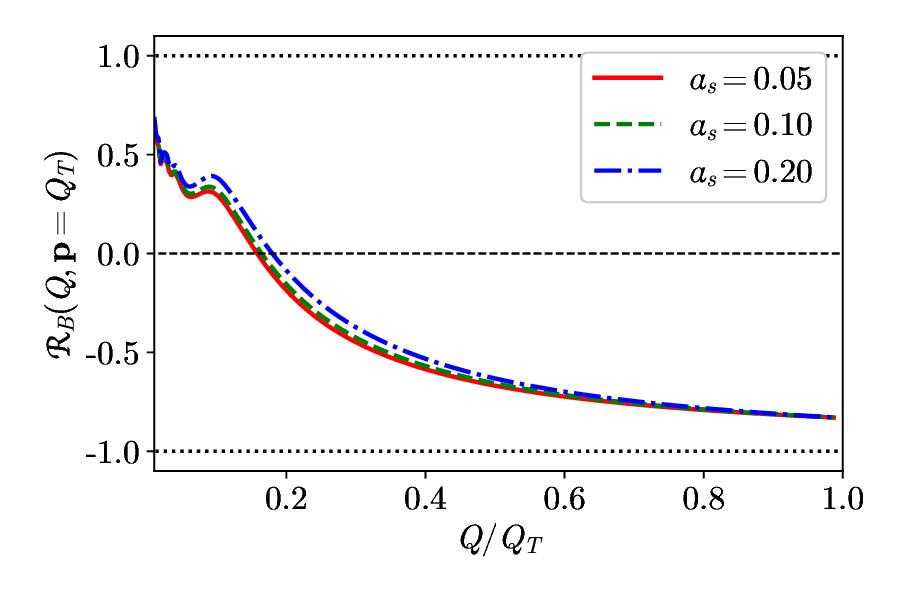}\includegraphics[width=0.5\textwidth]{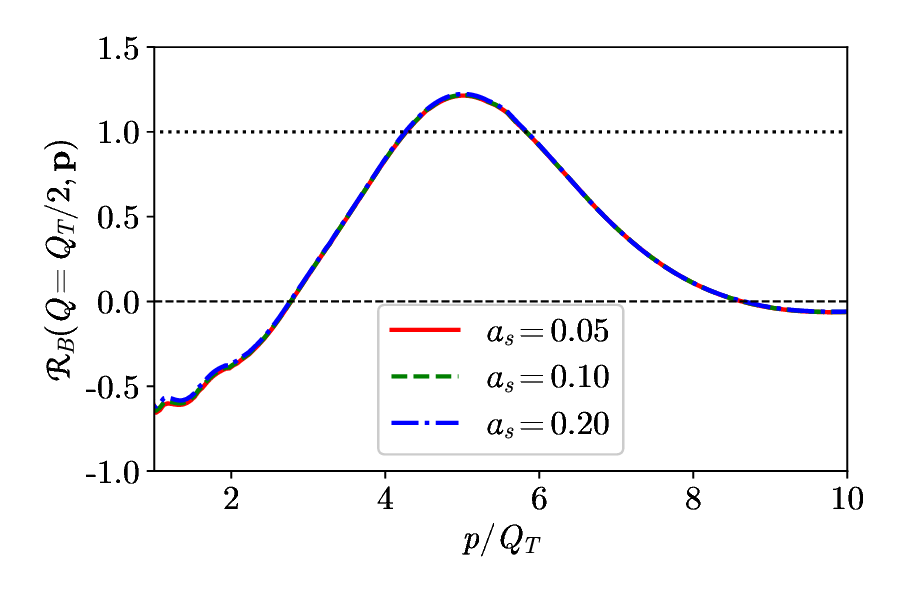}
\caption{$\mathcal{R}_B(Q,\mathbf{p})$ for $p=Q_T/10$ (top), $p=Q_T/2$ (middle) and $p=Q_T$ (bottom) versus $Q/Q_T$ (left plots), and for $Q=Q_T/10$ (top), $Q=Q_T/5$ (middle) and $Q=Q_T/2$ (bottom) versus $p/Q_T$ (right plots), for different values of $\alpha_s$, for $\cos\phi=0$.}
    \label{fig:RDeltaB}
\end{figure}

On Fig.\ref{fig:RDeltaB} we plot the unitarity ratio $R_B$.
The picture here is somewhat different. We again observe that generically this ratio is not small. However it almost vanishes for very small values of $p$ (the upper right plot) and is small for very large values of $p\gg Q_T$. This is almost the complementary region to that where $R_\Delta$ is small. It thus looks that the unitarity of $\mathbb{S}_Q$ is violated almost everywhere, but the nature of this violation is different in different kinematic regions. In some the main violation is due mostly to deviations of $\Delta_Q$ from $\Delta^U_Q$, while in others it is due to large deviations of $B_Q$ from $B^U_Q$.

We observe that the scaling of $R_B$ with $\alpha_s$ is not as good as for $R_\Delta$. Nevertheless it still holds in large kinematical regions.

\subsection{Scaling with $\alpha_s$}

Motivated by our numerical results that show a surprising scaling of the unitarity ratios with $\alpha_s$, here we attempt to understand the origin of this scaling. 

The first observation is that since both $R_\Delta$ and $R_B$ are ratios, the $\alpha_s$ independence is natural if perturbative corrections are small, and if the numerator and denominator in each ratio both are of order $\alpha_s$. Indeed one can formally expand these expressions to leading order.

The solution for $\lambda$ to first order in $\alpha_s$ is 
\begin{equation}
    \lambda_Q^c(\mathbf{p})=\left[1-\int_Q^{Q_T}{\frac{dQ^{\prime 2}}{Q^{\prime 2}}R(p,Q^\prime)}\right]\lambda_{Q_T}^c(\mathbf{p})+\mathcal{O}(\alpha_s^2).
\end{equation}
It then follows to first order
\begin{equation}
    \Delta_Q(\mathbf{p})=\left[1-\int_Q^{Q_T}{\frac{dQ'^2}{Q'^2}R_\Delta(p,Q')}\right]\Delta_{Q_T}(\mathbf{p})-\int_Q^{Q_T}{\frac{dQ'^2}{Q'^2}F(\mathbf{p},Q')}+\mathcal{O}(\alpha_s^2).
\end{equation}
So the denominator in $\mathcal{R}_\Delta(Q,\mathbf{p})$ reads
\begin{eqnarray}\label{eq:linden}
    \mathcal{D}^{\Delta}_Q(\mathbf{p})&\equiv&1-\frac{\Delta_Q(\mathbf{p})}{\Delta_{Q_T}(\mathbf{p})}=\frac{\alpha_s\beta^g_0}{4\pi}\left\{\int_Q^{Q_T}{\frac{dQ'^2}{Q'^2}\left[2J_0\left(\frac{p}{2Q'}\right)-1\right]}\right.\\
    &&+\left.\frac{1}{3\Delta_{Q_T}(\mathbf{p})}\int_Q^{Q_T}{\frac{dQ'^2}{Q'^2}\int{\frac{d^2\mathbf{k}}{(2\pi)^2}J_0(Q'^{-1}k)\lambda_{Q_T}^c(\mathbf{p}/2+\mathbf{k})\lambda_{Q_T}^c(\mathbf{p}/2-\mathbf{k})}}\right\}+\mathcal{O}(\alpha_s^2).\nonumber
\end{eqnarray}

On the other hand,
\begin{eqnarray}
\Delta_{Q}^U(\mathbf{p})&=&-\frac{1}{2}\int\frac{d^2\mathbf{k}}{(2\pi)^2}\left[1-\int_Q^{Q_T}{\frac{dQ^{\prime 2}}{Q^{\prime 2}}R(k,Q^\prime)}-\int_Q^{Q_T}{\frac{dQ^{\prime 2}}{Q^{\prime 2}}R(p-k,Q^\prime)}\right]\nonumber \\ &&\hskip 3cm \times\lambda_{Q_T}^c(\mathbf{k})\lambda_{Q_T}^c(\mathbf{p}-\mathbf{k})
+\mathcal{O}(\alpha_s^2),
\end{eqnarray}
and the numerator of $R_\Delta$ reads
\begin{eqnarray}
    \mathcal{N}^{\Delta}_Q(\mathbf{p})&\equiv&1-\frac{\Delta^U_Q(\mathbf{p})}{\Delta_{Q}(\mathbf{p})}=1
    +\frac{1}{2\Delta_{Q}(\mathbf{p})}\int \frac{d^2\mathbf{k}}{(2\pi)^2}\left(1-\int_Q^{Q_T}\frac{dQ^{\prime 2}}{Q^{\prime 2}}\left[R(k,Q^\prime)+R(p-k,Q^\prime)\right]\right)\nonumber \\
    &&\hskip 3cm \times  \lambda_{Q_T}^c(\mathbf{k})\lambda_{Q_T}^c(\mathbf{p}-\mathbf{k})+\mathcal{O}(\alpha_s^2)
\end{eqnarray}
which, expanding $1/\Delta_{Q}(\mathbf{p})$ to $\mathcal{O}(\alpha_s)$, results in
\begin{eqnarray}\label{eq:linnum}
    \mathcal{N}^{\Delta}_Q(\mathbf{p})&=&-\mathcal{D}^{\Delta}_Q(\mathbf{p})\\
   &&- \frac{\alpha_s\beta^g_0}{4\pi}\frac{1}{\Delta_{Q_T}(\mathbf{p})}\int{\frac{d^2\mathbf{k}}{(2\pi)^2}\int_Q^{Q_T}{\frac{dQ^{\prime 2}}{Q^{\prime 2}}\left[J_0\left(\frac{k}{2Q'}\right)-1\right]}\lambda_{Q_T}^c(\mathbf{k})\lambda_{Q_T}^c(\mathbf{p}-\mathbf{k})}
    +\mathcal{O}(\alpha_s^2).\nonumber
\end{eqnarray}

Finally,
\begin{equation}
    \mathcal{R}_\Delta(Q,\mathbf{p})=\frac{\mathcal{N}^{\Delta}_Q(\mathbf{p})}{\mathcal{D}^{\Delta}_Q(\mathbf{p})}\ ,
\end{equation}
with numerator and denominator given by~\eqref{eq:linnum} and~\eqref{eq:linden}, respectively. Since the expansion of both the numerator and denominator starts at order $\alpha_s$, the ratio to lowest order is independent of $\alpha_s$, but is nevertheless a nontrivial function of both $\bf p$ and $Q$.

The straightforward way to check that the scaling we observe is due the dominance of the first order term in $\mathcal{N}$ and $\mathcal{D}$ would be to numerically evaluate the expressions~\eqref{eq:linnum},~\eqref{eq:linden}. However, the 
numerical evaluation is rather tricky due to large cancellations that occur in the integral of the Bessel function $J_0$.

\begin{figure}[htp]
\centering
\includegraphics[width=0.5\textwidth]{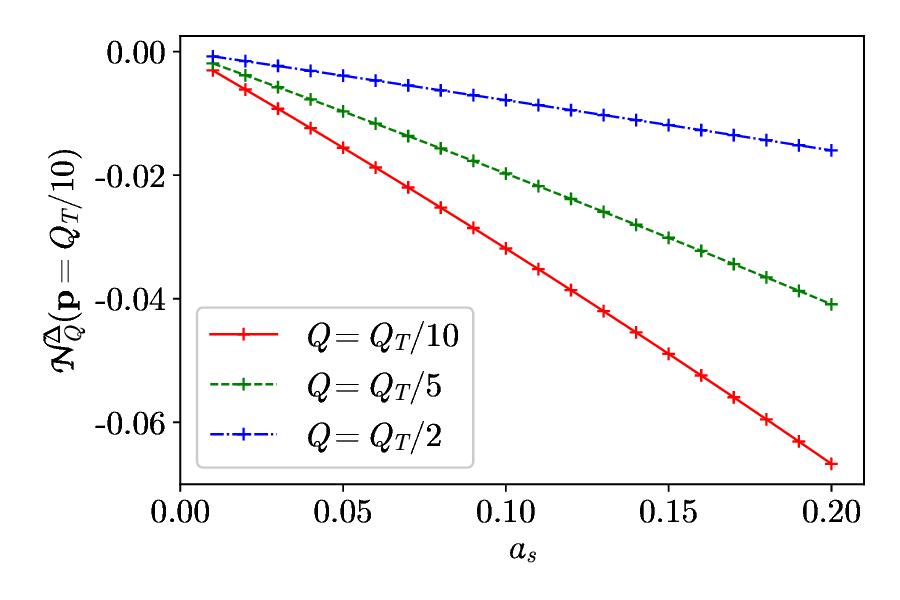}\includegraphics[width=0.5\textwidth]{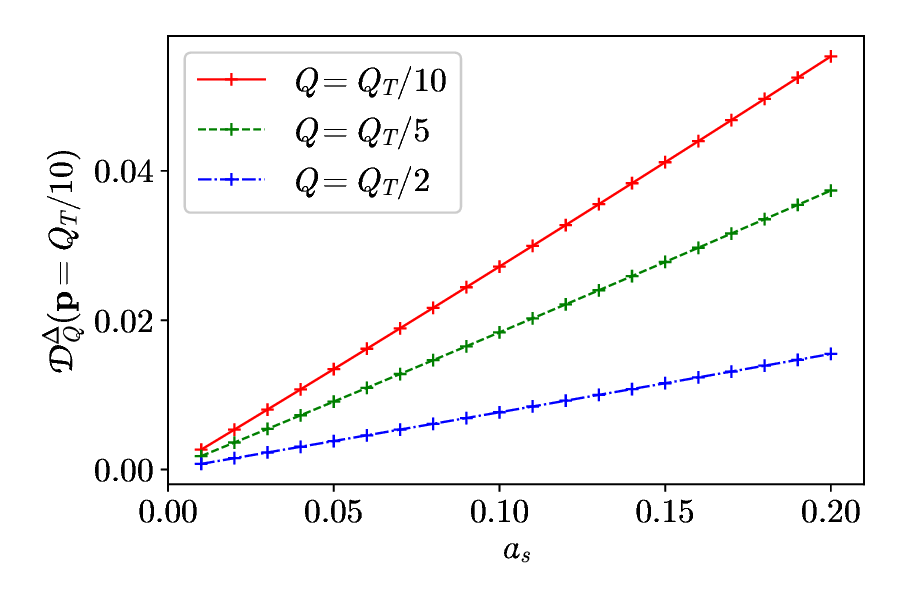}
\includegraphics[width=0.5\textwidth]{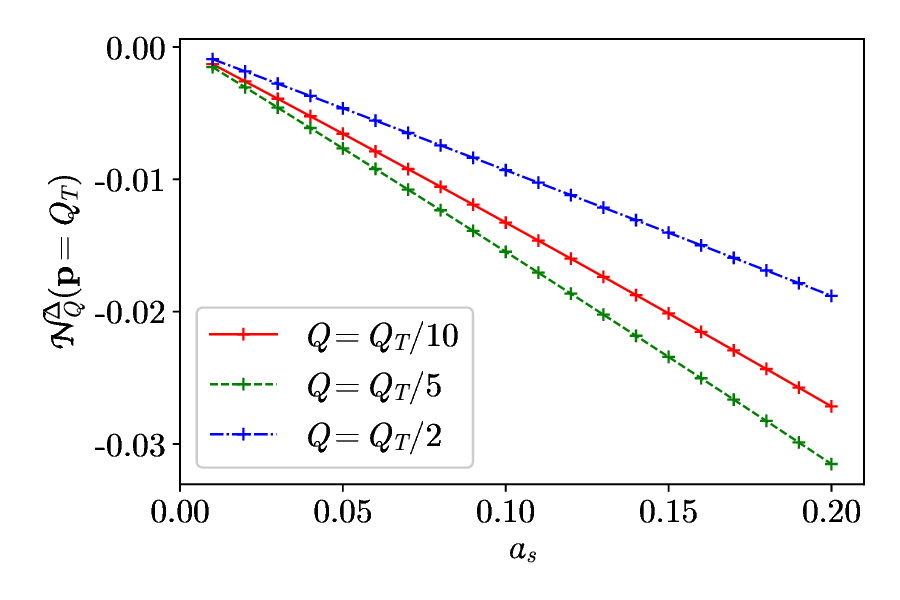}\includegraphics[width=0.5\textwidth]{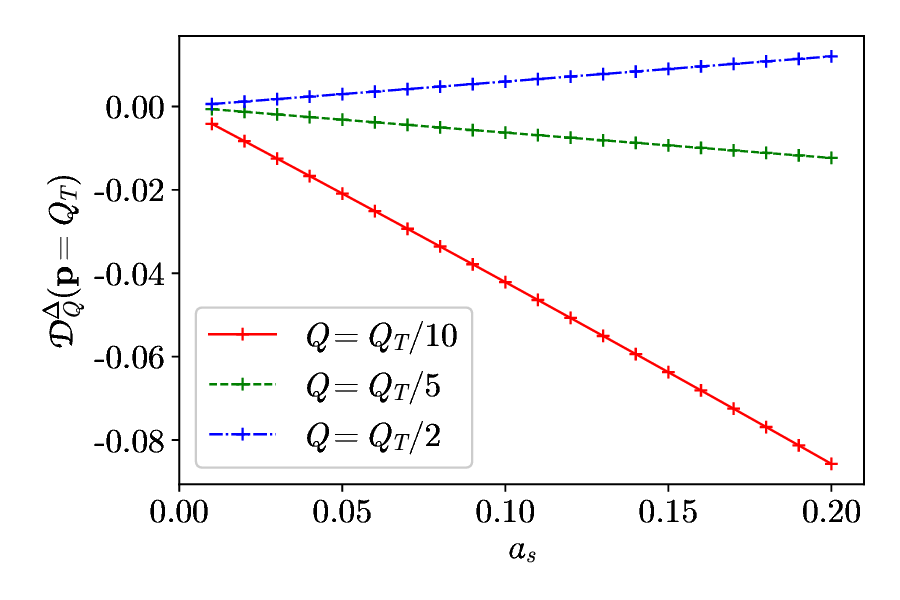}
\includegraphics[width=0.5\textwidth]{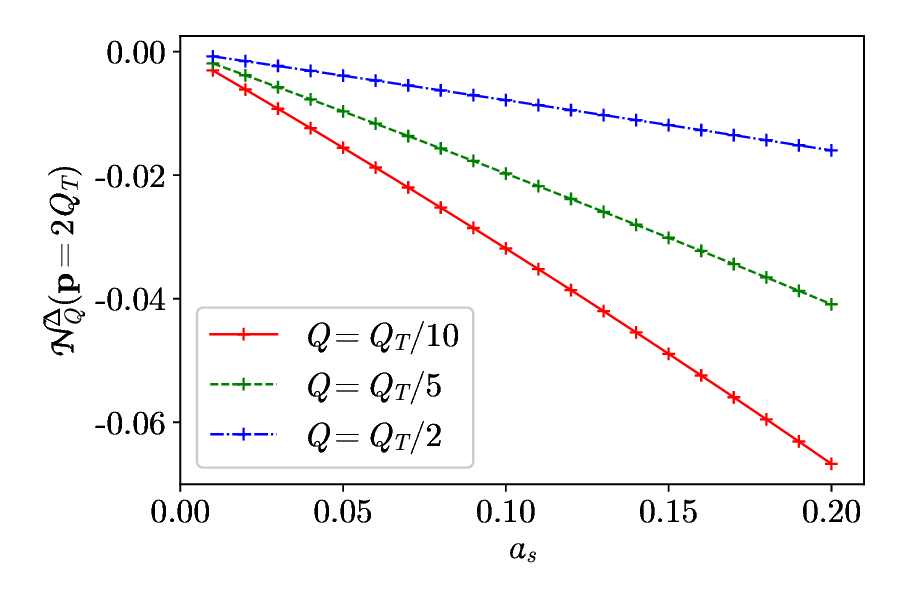}\includegraphics[width=0.5\textwidth]{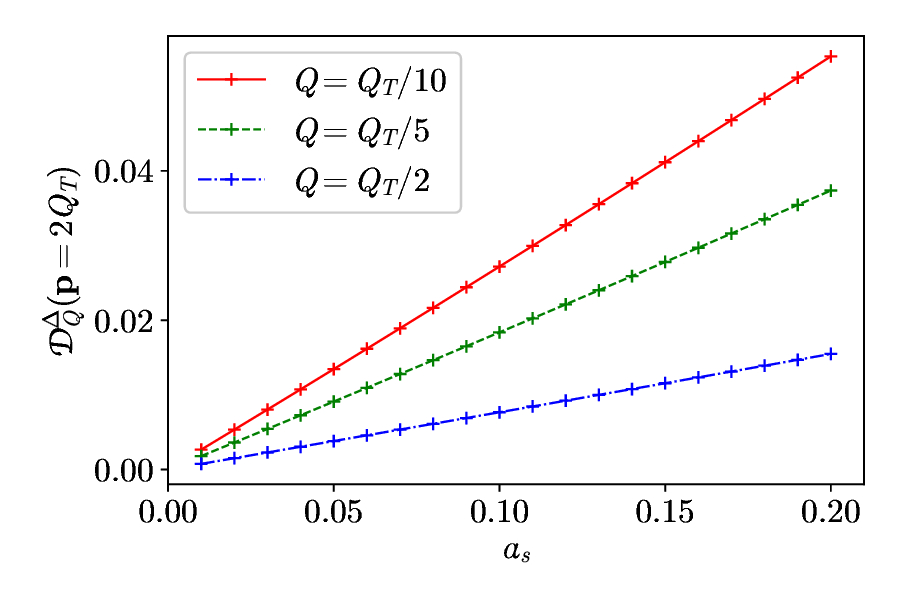}
\caption{Behavior of the numerator
(left plots) and denominator
(right plots) in \eqref{Rd}  versus $\alpha_s$ for $\mathbf{p}=Q_T/10$ (top), $Q_T$ (middle) and $2Q_T$ (bottom) for different values of $Q$ at $\phi=\pi/2$. }
    \label{fig:RDeltaQlin}
\end{figure}

Our strategy instead is to calculate directly the numerator and denominator in~\eqref{Rd} for different values of the coupling constant, and to check to what extent they are both linear functions of $\alpha_s$. On Fig.~\ref{fig:RDeltaQlin} we plot the numerator and the denominator in~\eqref{Rd} as a function of the QCD coupling constant $\alpha_s$ for different values of $p$ and $Q$. We observe a near perfect linear behavior for all the quantities plotted.

\begin{figure}[hbt]
\centering
\includegraphics[width=0.5\textwidth]{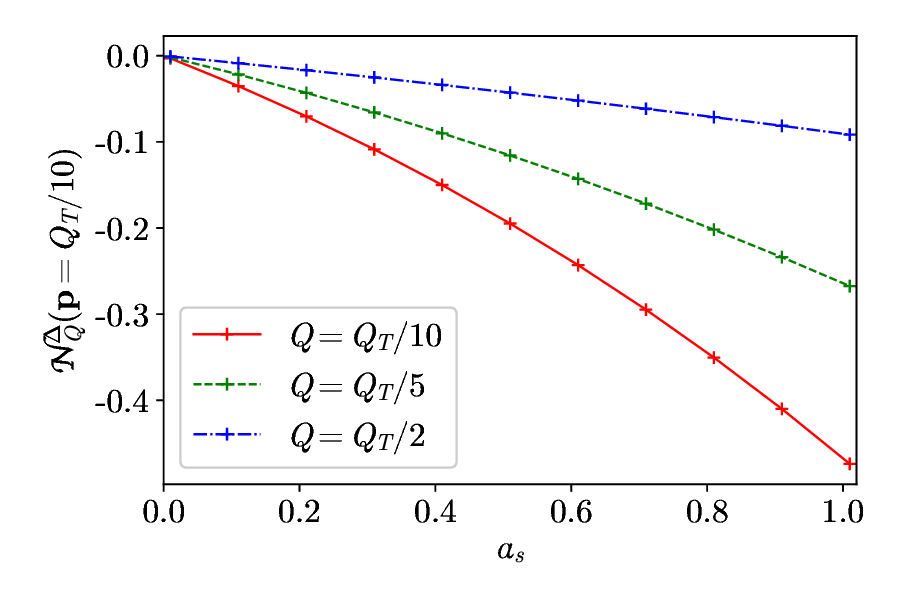}\includegraphics[width=0.5\textwidth]{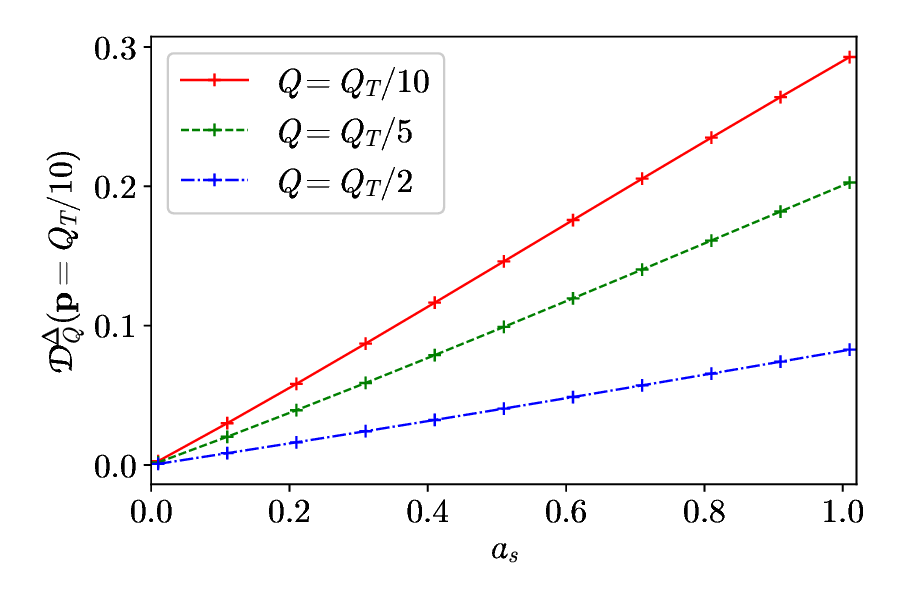}
\caption{Same as Fig.~\ref{fig:RDeltaQlin} for $\mathbf{p}=Q_T/10$ in a wider range of $\alpha_s$.}
\label{fig:RDeltaQlin2}
\end{figure}

On Fig.~\ref{fig:RDeltaQlin2} we plot the same for one value of $p$ but up to much higher values of $\alpha_s$. Surprisingly all the way up to $\alpha_s=1$ we observe the linear behavior of the denominator. The numerator starts to deviate from linearity for $\alpha_s\sim 0.3$ at small $Q$, but even then the deviation is rather modest.

It thus looks that the approximate independence of the ratios $R_\Delta$ on $\alpha_s$ is the reflection of the dominance of the $\mathcal{O}(\alpha_s)$ terms over higher order terms in perturbative expansion, in the numerator and denominator separately.
\section{Conclusions}
In this paper we studied the DGLAP resummation of the dressed gluon scattering amplitude discussed in~\cite{Kovner:2023vsy} necessary for taming large transverse logarithms on the NLO JIMWLK equation. For simplicity we considered the $SU(2)$ pure gauge theory. We limited ourselves to a weak field approximation, but kept higher order terms than those analyzed in ~\cite{Kovner:2023vsy} earlier.
We analyzed the evolution of the scattering matrix starting from the initial condition corresponding to a single dipole target.

Our main focus was on the question to what extent does the dressed gluon scattering matrix deviates from unitarity. Since this scattering matrix is truncated, in the sense that it does not include all possible final states, on general grounds we know that its unitarity is not required by the unitarity of the full $S$-matrix operator.

Indeed we find that the deviations of $\mathbb{S}_Q$ from unitarity are significant. In almost all kinematic ranges studied, this deviation is of the same order as the deviation of $\mathbb{S}_Q$ from the initial condition due to the evolution.

Remarkably we found that the ratios $\mathcal{R}_\Delta$~\eqref{eq:rdelta} and $\mathcal{R}_B$~\eqref{eq:rb},  sensitive to deviations from unitarity, have an extremely weak dependence on the QCD coupling constant. This is surprising, since we have studied the evolution down to rather small values of $Q$, i.e., $Q=Q_T/10$, and values of coupling constant up to $\alpha_s\sim 1$. One expects that the relevant parameter of the expansion is $\alpha_s\ln Q^2_T/Q^2$ which, in the range we probe, reaches values of order unity. 

It is an interesting question whether this feature persists for the $SU(3)$ gauge theory as well. One should also understand if this is an artifact of the weak field approximation we use, or that it survives in 
the full nonlinear regime as well. These questions are left for future study.

\acknowledgements
The research of AK was supported by the NSF Nuclear Theory grant \#2208387. This work is also supported by the U.S. Department of Energy, Office of Science, Office of Nuclear Physics, within the framework of the Saturated Glue (SURGE) Topical Theory Collaboration. AK thanks 
ITP at the University of Heidelberg  and CERN-TH group for support and hospitality.

The research of NA and VLP was supported by European Research Council project ERC-2018-ADG-835105 YoctoLHC, by Xunta de Galicia (CIGUS Network of Research Centres), by European Union ERDF, and by the Spanish Research State Agency under projects PID2020-119632GBI00 and PID2023-152762NB-I00. This work is part of the project CEX2023-001318-M financed by MCIN/AEI/10.13039/501100011033.

NA and VLP  thank Physics Department of the University of Connecticut for warm hospitality during the visit when this work was initiated.

\bibliography{mybib}

\end{document}